\RequirePackage{fix-cm}
\documentclass[natbib]{svjour3}                     
\smartqed  
\usepackage{graphicx}
\usepackage{epsf}
\usepackage{amssymb}
\usepackage{amsmath}
\usepackage{placeins}
\usepackage{color}

 \journalname{SSRv}
\newcommand{\be}{\begin{equation}}
\newcommand{\ee}{\end{equation}}
\newcommand{\beq}{\begin{eqnarray}}
\newcommand{\eeq}{\end{eqnarray}}
\newcommand\subsun[1]{{$_{\normalsize\odot}$}}

\newcommand{\kms}{~km~s$^{-1}$}

\def\lsim{\;\raise0.3ex\hbox{$<$\kern-0.75em\raise-1.1ex\hbox{$\sim$}}\;}
\def\gsim{\;\raise0.3ex\hbox{$>$\kern-0.75em\raise-1.1ex\hbox{$\sim$}}\;}

\def\cmc{~cm$^{-3}$}

\def\arcmin{\hbox{$^\prime$}}

\begin{document}
\title{Structures and components in galaxy clusters: observations and
models}

\titlerunning{Structures and components in clusters}        

\author{A.M.~Bykov \and
        E.M.~Churazov \and
        C.~Ferrari  \and
        W.R.~Forman \and
        J.S.~Kaastra \and
        U.~Klein \and
        M.~Markevitch \and
        J.~de Plaa
}


\institute{A.M. Bykov \at
              Ioffe Institute, 194021, St. Petersburg, Russia\\ 
              \email{byk@astro.ioffe.ru}
            \and
           E.M. Churazov \at
			Max Planck Institute for Astrophysics, Karl-Schwarzschild-Str. 1, 85741
			Garching, Germany;\\
			Space Research Institute (IKI), Profsoyuznaya 84/32, Moscow 117997,
			Russia\\
              \email{churazov@mpa-garching.mpg.de}
              \and
              C. Ferrari \at
              Laboratoire Lagrange, UMR7293, Universit\'e de Nice Sophia-Antipolis, CNRS, Observatoire de la C\^ote d'Azur \\
              \email{chiara.ferrari@oca.eu}
            \and
           W.R. Forman \at
            Harvard Smithsonian Center for Astrophysics, 60 Garden Street, Cambridge, MA 02138, USA \\
               \email{forman@cfa.harvard.edu}
	       \and
           J.S.  Kaastra \at
           SRON Netherlands Institute for Space Research, Sorbonnelaan 2,
	   3584 CA Utrecht, The Netherlands \\
                \email{j.kaastra@sron.nl}
            \and
           U. Klein \at
           Argelander-Institut f\"ur Astronomie, University of Bonn, Germany\\
                \email{uklein@astro.uni-bonn.de}
            \and
           M. Markevitch \at
           Astrophysics Science Division, NASA/Goddard Space Flight Center, Greenbelt, MD 20771, USA\\
                \email{maxim.markevitch@nasa.gov}
            \and
           J. de Plaa  \at
            SRON Netherlands Institute for Space Research, Sorbonnelaan 2,
	   3584 CA Utrecht, The Netherlands \\
                \email{J.de.Plaa@sron.nl}
                }

\date{Received: date / Accepted: date}

\maketitle

\begin{abstract}
Clusters of galaxies are the largest gravitationally bounded structures in the Universe dominated by dark matter. We review the observational appearance and physical models of plasma structures in clusters of galaxies. Bubbles of relativistic plasma which are  inflated by supermassive black holes of AGNs, cooling and heating of the gas, large scale plasma shocks, cold fronts,  non-thermal halos and relics are observed in clusters. These constituents are reflecting both the formation history and the dynamical properties of clusters of galaxies. We discuss X-ray spectroscopy as a tool to study the metal enrichment in clusters and  fine spectroscopy of Fe X-ray lines as a powerful  diagnostics of both the turbulent plasma motions and the energetics of the non-thermal electron populations. The knowledge of the complex dynamical and feedback processes is necessary to understand the energy and matter balance as well as to constrain the role of the non-thermal components of clusters.
\end{abstract}

\section{Introduction}
\label{sec:1}
Clusters of galaxies grow by gravitational collapse to the most
massive objects of the Universe. While the total mass is dominated by
dark matter ($\sim 80$\%), there is a significant baryonic
contribution of 20\% and the processes where baryons are involved
prominently determine the evolutionary physics and the observational
appearance of clusters.

In this paper we focus on the baryonic component of clusters and in
particular on the hot gas that, with its thermal and non-thermal
constituents, comprises the majority of these baryons. Cluster
galaxies are embedded in this hot intracluster medium (ICM), but
represent a small fraction of both the volume and the baryonic mass.

For most clusters, the hot gas reaches temperatures of
$10^7$--$10^8$~K. At the low temperature end, below temperatures of
about $2\times 10^7$~K, it is more common to speak about groups of
galaxies rather than clusters of galaxies, but the transition is of
course smooth.

From an X-ray perspective, clusters are found in two variants: those
with a cool core and those lacking such a core. In the cool core
clusters the density of the gas in the center reaches values of $\sim
10^{-2}$~cm$^{-3}$ (that is, for instance, only 1\% of the typical
density of the interstellar medium of a galaxy). The density decreases
rapidly towards the outskirts down to levels of order
$10^{-4}$~cm$^{-3}$ or less.

The density in the cool cores is high enough to cause significant
cooling over cosmological time scales through thermal bremsstrahlung
emission observed in X-rays. Without any heating mechanism, this gas
would cool down further and do form stars. However, it appears that
active nuclei in the core of the dominant cluster galaxies can emit so
much power that the associated heating compensates the cooling of the
central gas. This leads to a physically interesting but complex
feedback loop between the central supermassive black hole and the
cluster gas.  These processes and the associated plasma structures
will be discussed in this review.

Clusters are by no means static entities, they still grow. Large scale
violent processes occur, like cluster mergers or, more frequently, the
capture of groups or individual galaxies by massive systems. Due to
the supersonic velocities involved, shocks are produced at various
locations within colliding clusters. This leads to local heating,
particle acceleration and modification of the magnetic fields, and we
will also discuss these processes.

Another part of the bulk motions may cascade downwards in scale in the
form of turbulence, that represents a contribution to the total
thermal pressure of the order of 5--15\% and is stronger in the outer
parts of clusters \citep[see for review][]{2012ARA&A..50..353K,dolag08}. In
these external regions, infalling galaxies and groups will give rise
to density inhomogeneities which, due to the relatively low density in
cluster outskirts, take a long time to disappear.

Mixing of merging components also leads to interesting processes at
the interface between cold and hot gas. Further, when galaxies are
being captured by a cluster, they may lose their chemically enriched
gas to the intracluster medium by ram pressure or interaction with
other galaxies. The ICM itself is a rich archive of the past chemical
history of the cluster: it contains information on the distribution of
stars and the relative frequencies of specific subclasses.

Clusters, that are the largest gravitationally bound systems in the
Universe, form a rich and living laboratory to study all kinds of
processes that shape their appearance. Our review is devoted to
discussion of plasma structures of different scales and origin in
clusters of galaxies. More general discussion and many important ideas
about the evolution and cosmological importance of clusters of
galaxies can be found in recent reviews \citep[see e.g.][and
references
therein]{boehringer10,allen11,2012ARA&A..50..353K,planelles14}.

\section{AGNs in galaxy clusters. Feedback processes.}
\label{sec:3.1.1}

The gas in the cores of galaxy clusters has a radiative cooling time
of the order of $10^9$ years or less, opening the possibility of
forming an extremely massive central galaxy. This occurs in numerical
simulations with radiative cooling, but is generally not observed in
nature, with a few observed exceptions, such as, e.g., the Phoenix
cluster \citep{2012Natur.488..349M}. Instead, observations suggest
that the mechanical energy released by a central AGN regulates the
thermal state of the gas, preventing it from catastrophic
cooling. Bubbles of relativistic plasma are inflated by a supermassive
black hole and rise buoyantly through the gaseous atmosphere, leading
to a number of spectacular phenomena such as expanding shocks, X-ray
dim and radio bright cavities, old and ``dead'' cavities and filaments
in the wakes of the rising bubbles formed by entrained low entropy
gas.

Simple arguments based on the energy content of bubbles and their
life-time show that the amount of mechanical energy supplied by AGNs
matches approximately the gas cooling losses in objects vastly
different in size and luminosity. This hints at some form of
self-regulation of the AGN power. How the mechanical energy, provided
by the AGN, is dissipated depends on the ICM microphysics (e.g.
magnetic fields, viscosity, conduction etc). However it is plausible
that close to 100\% of the mechanical energy is eventually dissipated
in the cluster core, regardless of the particular physical process
involved.

AGN feedback is plausibly a key process for the formation of massive
ellipticals at $z\sim 2-3$, as suggested by the correlation of galaxy
bulge properties and the mass of the SMBH
\citep{2000ApJ...539L...9F,2000ApJ...539L..13G}. Galaxy clusters offer
us a zoomed view on this process at $z\sim 0$. Three pre-requisites
are needed for this scenario to work (i) a hot gaseous atmosphere in
the galaxy is present, (ii) the black hole is sufficiently massive and
(iii) a large fraction of AGN energy is in mechanical form. The latter
depends critically on the physics of accretion, in particular, on the
transition of the SMBH energy output from the radiation-dominated mode
to the mechanically-dominated mode when the accretion rate drops below
a fraction of the Eddington value
\citep[e.g.,][]{2005MNRAS.363L..91C}. Given that the coupling constant
of these two forms of energy output with the ICM can differ by a
factor of $10^{4}-10^{5}$, this change in the accretion mode may
explain the switch of a SMBH (and its parent galaxy) from the QSO-type
behavior and an intense star formation epoch to the radiatively
inefficient AGN and essentially passive evolution of the parent
galaxy.

Below we briefly outline only most general features of the AGN
Feedback model in galaxy clusters and do not discuss results from
numerical simulations.  Extended reviews on the AGN Feedback in
clusters can be found in, e.g.,
\citet{2007ARA&A..45..117M,2012ARA&A..50..455F}.

\subsection{Cluster cores without AGN feedback}

The radiative cooling time of the gas in the central parts of rich
galaxy clusters ($\displaystyle t_{\rm
  cool}=\frac{\gamma}{\gamma-1}\frac{nkT}{n^2 \Lambda(T)}\sim {\rm
  ~few~} 10^{8}-10^{9}~{\rm yr}$) is shorter than the Hubble time
\citep[e.g.,][]{1976ApJ...203..569L,1977ApJ...215..723C,1977MNRAS.180..479F}.
Here $n$ and $T$ are the density and temperature of the ICM,
respectively, $\gamma$ is the adiabatic index and $\Lambda(T)$ is the
radiative cooling function.  Since $\displaystyle t_{\rm cool}\propto
1/n$, the cooling time is short in the center, but rapidly increases
with radius. The radius $r_{\rm cool}$ where $t_{\rm cool}\sim
t_{Hubble}$ is usually referred to as the cooling radius. Without an
external source of energy, the gas inside $r_{\rm cool}$ must cool and
flow towards the center, forming a so-called ``cooling flow''
\citep[see][for a review of the scenario without AGN
feedback]{1994ARA&A..32..277F}. Observations
\citep[e.g.,][]{2006PhR...427....1P,2001ApJ...557..546D}, however,
suggest that the net rate of gas cooling to low temperatures is a
small fraction ($\sim$10\% or below) of the straightforward estimate:

\begin{equation}
\dot{M}_{\rm cool}=\frac{L_{\rm cool}}{\frac{\gamma}{\gamma-1}kT}\mu m_{\rm p}\sim
10^{2}-10^{3} M_\odot~{\rm yr}^{-1},
\end{equation}

\noindent where $L_{\rm cool}$ is the total cooling rate within
$r_{\rm cool}$, $\mu$ is the mean particle atomic weight. This implies
that a source of heat is needed to offset ICM cooling losses. In the
late 90's and early 2000's it became clear that a supermassive black
hole sitting at the center of the dominant cluster galaxy could
operate as such source.

The above ``cooling flow'' problem applies not only to rich clusters
and groups, but also to individual hot gas rich galaxies
\citep[e.g.,][]{1986MNRAS.222..655T}.  The fact that these galaxies do
not show significant star formation argues for feedback from these low
mass systems (some with SMBH masses as large as those in central
galaxies in rich clusters) up to the most massive and X-ray luminous
clusters.

\subsection{Evidence for AGN mechanical feedback and its energetics}

Massive ellipticals at the cores of rich clusters host black holes
with masses larger than $10^9~M_\odot$. In terms of energetics, such
black holes, accreting at the Eddington rate, could release up to
$10^{47}~{\rm erg~s^{-1}}$ -- much more than needed to reheat the
gas. However, we do not find extremely bright AGNs in nearby clusters,
and the coupling of radiation to the fully ionized ICM (via Compton
scattering) is weak. Based on radio observations,
\citet{1990MNRAS.246..477P} argued that mechanical power of jets in
NGC\,1275 (dominant galaxy in the Perseus cluster) is likely much
higher than their radiative power and it might be comparable to the
gas cooling losses. For less massive systems
\citet{1993MNRAS.263..323T} and \citet{1995MNRAS.276..663B} suggested
that the mechanical power of jets may have a strong impact on the
thermal state of the gas in the central region. But it was only the
combination of X-ray and radio data that convincingly demonstrated a
crucial role of AGN feedback in cluster cores.

By now, Chandra and XMM-Newton found signs of AGN/ICM interactions in
a large fraction of relaxed clusters
\citep[e.g.,][]{2000ApJ...534L.135M,2006MNRAS.373..959D,2006ApJ...652..216R,2012MNRAS.421.1360H,2012MNRAS.427.3468B}.
But clear evidence of this process in the Perseus cluster and in
M87/Virgo had previously been seen in ROSAT images
\citep{1993MNRAS.264L..25B,1995MNRAS.274L..67B}. Based on these
sketchy X-ray and radio images, the basic features of the mechanical
feedback model and its energetics, inspired by analogy with powerful
atmospheric explosions, were outlined \citep{2000A&A...356..788C} just
before Chandra and XMM-Newton launch.

This impact of the SMBH mechanical power on the ICM can be best
illustrated for M87 - the X-ray brightest elliptical galaxy in the
nearby Virgo cluster.  While M87/Virgo is not a very luminous system
($L_{\rm cool}$ is of the order $10^{43}$~ erg\,s$^{-1}$, e.g.
\citealt{1998MNRAS.298..416P,2000ApJ...543..611O}), its proximity
(distance $\sim$ 16 Mpc) offers an exquisitely detailed view on the
processes in the very core of the galaxy
\citep[e.g.][]{2005ApJ...635..894F,2007ApJ...665.1057F,2010MNRAS.407.2046M,2010MNRAS.407.2063W}.

\begin{figure*}
\includegraphics[width=0.99\columnwidth]{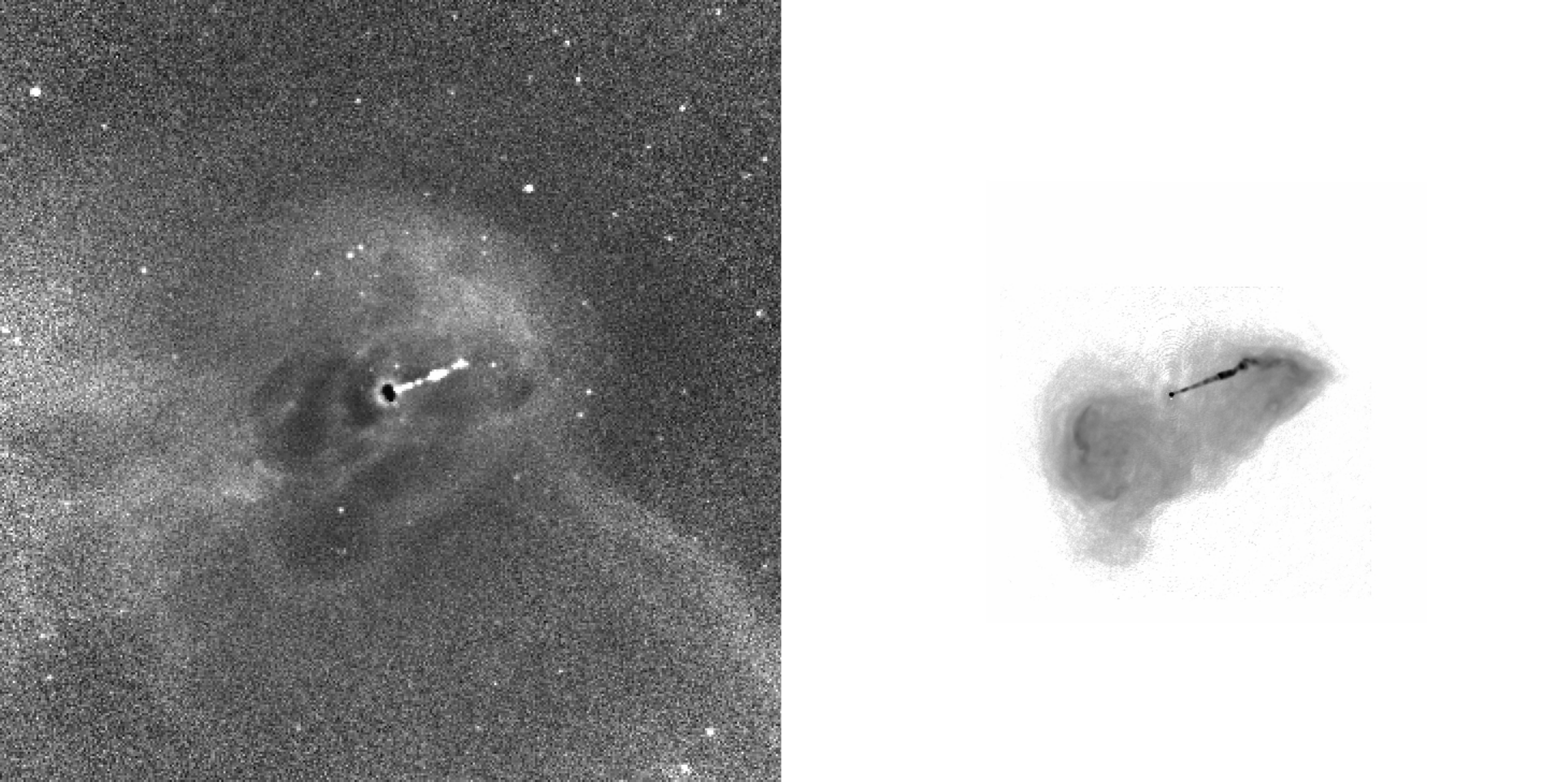}
\caption{X-ray \citep{2007ApJ...665.1057F} and 6 cm radio
\citep{1989ApJ...347..713H} images of the M87 core ($3'\times3'$). The X-ray
image (left) was flat-fielded to emphasize various non-axisymmetric features.
Images are shown to scale, centered at the supermassive black hole with a kpc
scale jet going to the NW. X-ray cavities,  matching the 6 cm image of the inner
radio lobes are clearly seen in the left image.}
\label{fig:m87x6}
\end{figure*}

Shown in Fig.~\ref{fig:m87x6} are the X-ray and 6~cm radio images of
M87 (central $3'\times3'$). Synchrotron emission for a jet coming from
the SMBH is clearly visible. Except for the jet, the X-ray image is
dominated by the thermal emission of the optically thin plasma at a
temperature 1-2 keV (unambiguously proven by the observed
spectra). The radio image on the right shows synchrotron emission from
the jet and radio lobes filled with relativistic plasma. The
radio-bright lobes at 6 cm nicely correspond to the X-ray dim regions
that are defined by X-ray bright shells, suggesting that thermal
plasma is displaced by the relativistic plasma within the central $1'$
from the core. A minimum energy, required to (slowly) inflate a bubble
of a given volume at a constant pressure is the enthalpy:

\begin{equation}
E_{\rm bub}=\frac{\gamma}{\gamma-1}PV,
\label{eq:ebub}
\end{equation}

\noindent where $\gamma$ is the adiabatic index of the fluid inside
the bubble ($\gamma=4/3$ and $5/3$ for mono-atomic relativistic and
non-relativistic fluids respectively), $P$ is the ICM pressure and $V$
is the volume. This gives a lower limit on the amount of mechanical
energy produced by the SMBH. Since we do not see evidence for a very
strong shock surrounding the lobes (see, however, below), the true
amount of energy should not be far from this lower limit.

To evaluate the mechanical power of the AGN, we need to estimate the
life-time of the bubbles $t_{\rm bub}$. The simplest recipe
\citep{2000A&A...356..788C} comes from the analysis of the buoyancy
driven evolution of the bubble. The importance of buoyancy for the
radio lobes, filled with relativistic plasma, was pointed out by
\citet{1973Natur.244...80G}. Comparing the expansion velocity (which
depends on the AGN power and the size of the bubble) and the buoyant
rise time (which depends on the gravity and the size of the bubble)
one gets an upper limit on the life time of the bubble. For a steady
AGN power $L_{\rm AGN}$ deposited into a small volume, the initial
expansion of the bubble is supersonic, but it slows as the bubble
grows.  Soon after the expansion velocity becomes subsonic, the bubble
is deformed by the Rayleigh-Taylor instability and rises under the
action of buoyancy. The terminal velocity of the rising bubble $v_{\rm
  rise}$ is set by the balance of the ram pressure (assuming low
viscosity of the ICM)and the buoyancy force:

\begin{equation}
g\frac{4}{3}\pi r^3\rho_{\rm gas}\approx C \pi r^2 \rho_{\rm gas} v^2,
\end{equation}

\noindent where $g$ is the gravitational acceleration, $r$ is the
bubble radius, $\rho_{\rm gas}$ is the ICM density and $C$ is a
dimensionless constant of order unity. Thus $\displaystyle v_{\rm
  rise}\sim \sqrt{gr}$. At the same, time AGN activity drives the
expansion of the bubble with the velocity $v_{\rm exp}$ set by the AGN
power $L_{\rm AGN}$, e.g., from Eq. (\ref{eq:ebub}) {\bf
  $\displaystyle v_{\rm exp}\sim \frac{\gamma-1}{\gamma}\frac{L_{\rm
      AGN}}{4 \pi P r^2}$}, provided that the expansion velocity is
subsonic. The condition $v_{\rm exp} \gtrsim v_{\rm rise}$ sets the
lower limit on the AGN power needed to ignore the role of
buoyancy. For M87 the size of the inner lobes suggests the jet
mechanical power $\displaystyle L_{\rm AGN}\sim \frac{\gamma}{\gamma
  -1}\frac{4\pi Pr^2}{\sqrt{gr}}\sim$ few $10^{43}$~erg\,s$^{-1}$. Of
course such estimates (and various modifications e.g.,
\citealt{2007ARA&A..45..117M}) are only accurate to within a factor of
a few. Nevertheless, it is clear that mechanical input of the SMBH is
sufficient to offset gas cooling losses, i.e.  $L_{\rm AGN}\sim
10^{43}$~erg\,s$^{-1} \sim L_{\rm cool}$.

\begin{figure*}
\includegraphics[width=0.5\columnwidth, bb = 19 -30 575 504]{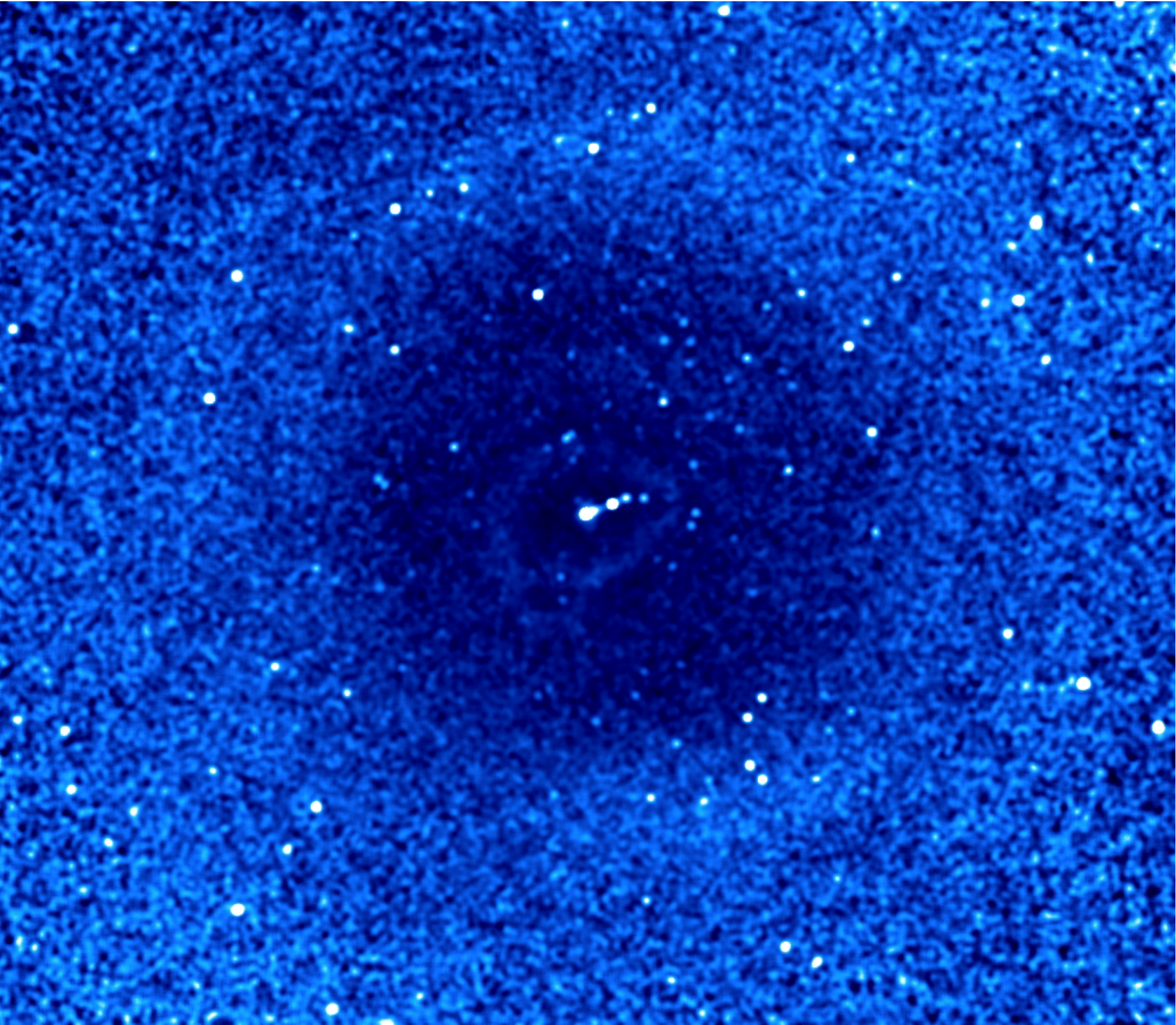}
\hfill
\includegraphics[width=0.5\columnwidth, bb = 18 30 575 504]{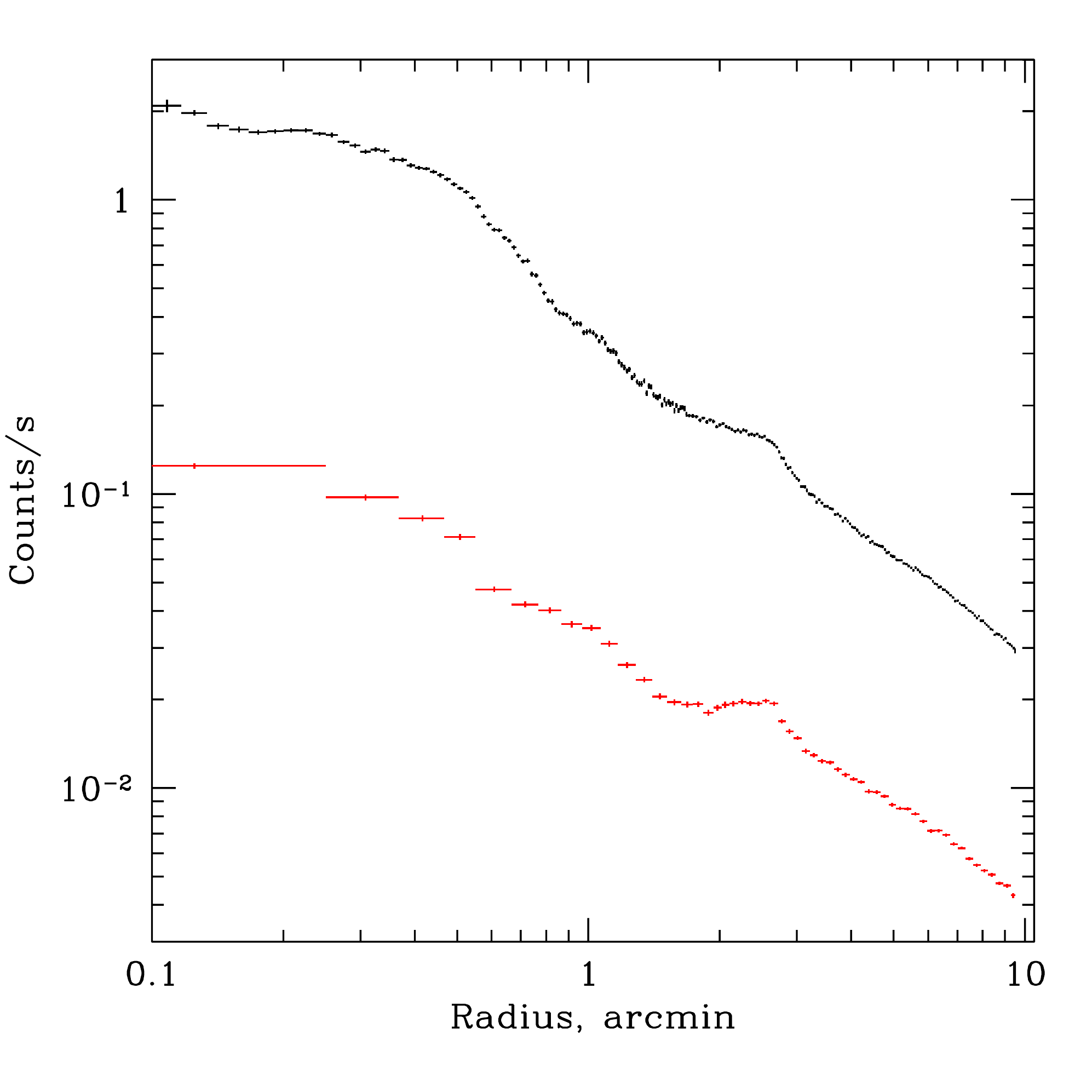}
\caption{{\bf Left:} Chandra image of M87 ($\sim 7'\times 7'$) in the 3.5--7.5
keV energy band divided by a spherically symmetric model. This energy band shows
pressure variations in the gas \citep{2007ApJ...665.1057F}. A nearly perfect
ring at $\sim 2.75'$ (12.8 kpc) is clearly seen. This is a characteristic
signature of a shock, driven by an outburst from the central SMBH. {\bf Right: }
The surface brightness profiles \citep{2007ApJ...665.1057F} in the 1.2--2.5 keV
(upper curve) and 3.5--7.5 keV (lower curve) bands show a prominent feature at
$2-3'$, along with a fainter feature at $0.6'$.}
\label{fig:m87shock}
\end{figure*}

One can expect that the initial phase of the bubble expansion is
supersonic and it will drive a shock into the ICM.  As the expansion
slows, the shock weakens and moves ahead of the expanding boundary.
These shocks offer yet another way to measure the AGN power. In M87 we
see at least two generations of shocks
\citep{2005ApJ...635..894F,2007ApJ...665.1057F} at $0.6'$ and $2.7'$
from the center (Fig.~\ref{fig:m87shock}).  The hard emission
(3.5--7.5 keV) shows a ring of emission with an outer radius ranging
from $2.5'$ to $2.85'$ (11.6--13.3 kpc).  This ring of hard emission
provides an unambiguous signature of a weak shock. The gas temperature
in the ring rises from $\sim2.0$ keV to $\sim2.4$ keV implying a Mach
number of ${\cal M} \sim 1.2$ (shock velocity $v = 880$~km\,s$^{-1}$ for a 2
keV thermal gas). At the shock, the density jump is 1.33 which yields
a Mach number of 1.22, consistent with that derived from the
temperature jump. The total energy needed to drive this shock can be
readily estimated from 1D hydro simulations $E\sim 5\times
10^{57}$~erg. The age of the outburst that gave rise to the shock is
$\sim 12$~Myr. If we divide the energy of the outburst by the age of
the shock we get a mean energy release $\sim 10^{43}$~erg\,s$^{-1}$ --
broadly in agreement with the ICM cooling losses.

\begin{figure*}
\includegraphics[width=0.99\columnwidth]{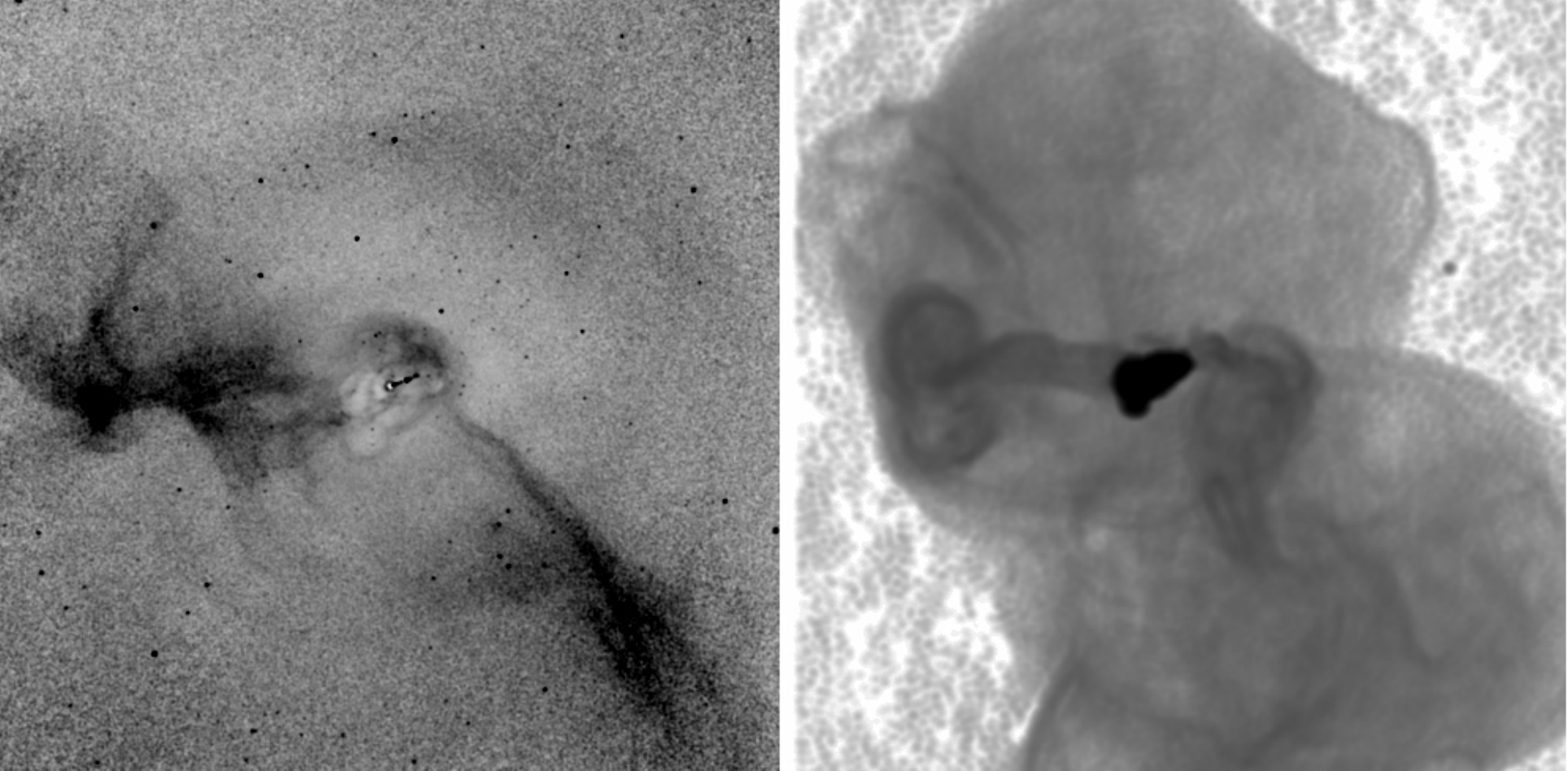}
\caption{X-ray \citep{2007ApJ...665.1057F} and 90 cm radio
\citep{2000ApJ...543..611O} images (right) of the core of M87 ($8'\times8'$).
Filaments of cool gas are entrained by the buoyantly rising bubbles.}
\label{fig:m87x90}
\end{figure*}

\begin{figure*}
\includegraphics[width=0.99\columnwidth]{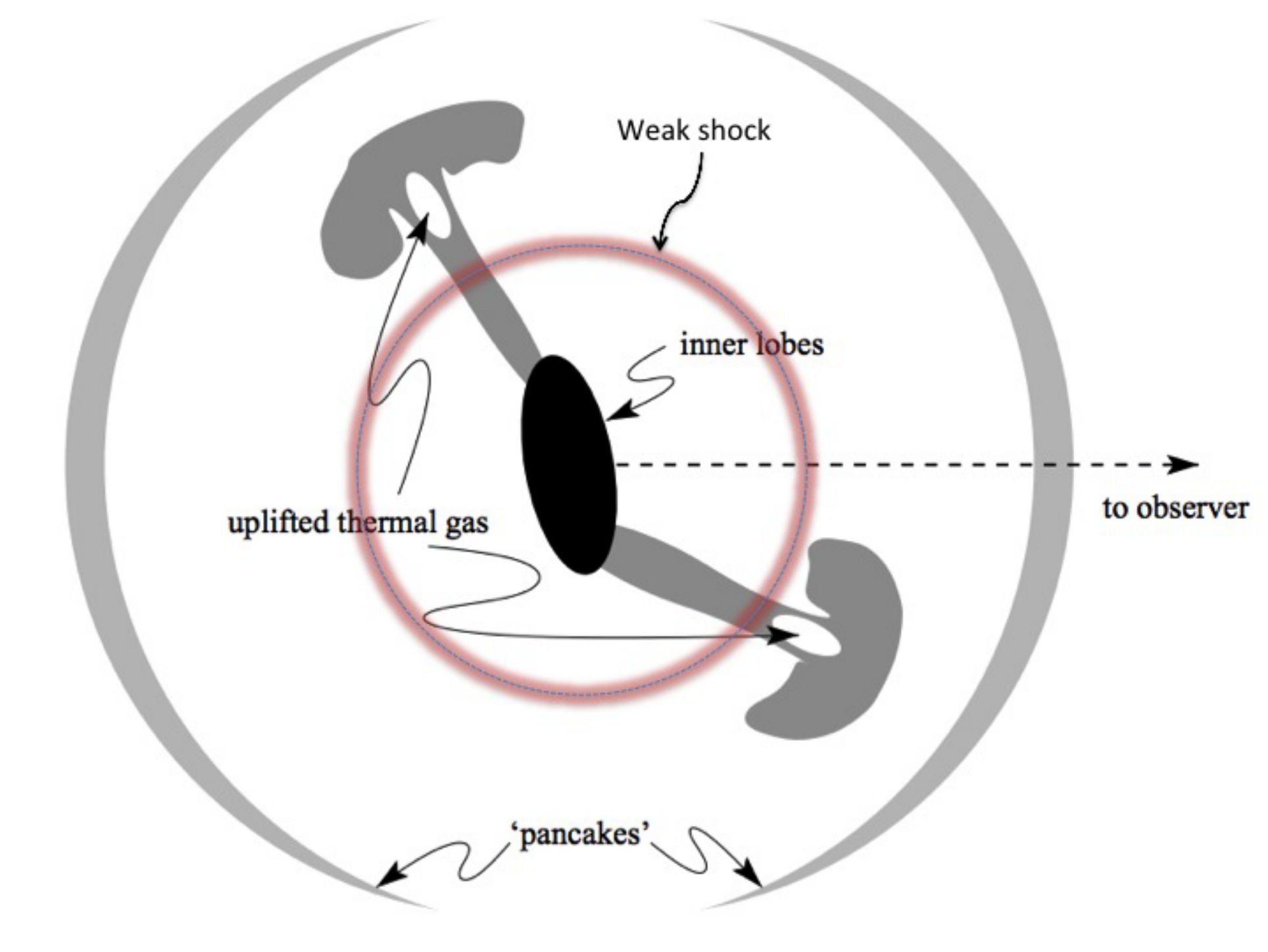}
\caption{Schematic picture of major signs of AGN/ICM interaction \citep[adapted
and modified from][]{2001ApJ...554..261C}, inspired by analogy with mushroom
clouds produced by powerful atmospheric explosions. The black region in the
center denotes the inner radio lobes, driven by the SMBH mechanical power. The
circular structure is a weak shock wave produced by these inner lobes. Gray
``mushrooms'' correspond to the buoyant bubbles already transformed into tori,
and the gray lens-shaped structures are the pancakes formed by the older bubbles
(c.f. Fig.~\ref{fig:m87x90}).}
\label{fig:m87_sketch}
\end{figure*}

\subsection{Dissipation of mechanical energy}

Once the bubble is detached from the central source, its evolution is
governed by buoyancy. During the rise the bubble may transform into a
toroidal structure as the``ear-like'' structure in M87 (see
Fig.~\ref{fig:m87x90}) which resembles a mushroom formed by a powerful
atmospheric explosion.  As in case of the atmospheric explosion, the
bubble is able to entrain large amounts of ambient gas from the core
of cluster and transport it to large distance from the cluster center
\citep[e.g.,][]{2001ApJ...554..261C,2003MNRAS.344L..48F,2010MNRAS.407.2063W}.

Note that adiabatic expansion of rising bubbles leads to a rapid
decrease of the radio emission, since both the magnetic field
strength, and the density and energy of relativistic particles are
decreasing at the same time. This decrease is especially strong if the
aging break in the distribution of electrons is brought by adiabatic
expansion into the observable frequency range. Thus, unless there is
continuous reacceleration of electrons, the radio bright bubbles
should evolve into a radio dim objects. Since the pressure support
inside the bubble could still come from magnetic fields and low energy
electrons and protons (Lorentz factor of 1000 or lower), the bubble is
still seen as an X-ray cavity, but is very dim in radio. Such ghost
bubbles are believed to be widespread in the cluster cores and one can
expect many of them to be detected with a new generation of a low
frequency instruments \citep{2002A&A...384L..27E}.

The bubble rise velocity $v_{\rm rise}$ is smaller than the ICM sound
speed, and it is much smaller than the sound speed of the relativistic
fluid inside the bubble. Adiabatic expansion of the rising bubble
implies that its enthalpy is decreasing according to the ambient gas
pressure $\displaystyle H=\frac{\gamma}{\gamma-1}PV\propto
P^{\frac{\gamma-1}{\gamma}}$. This means that after crossing a few
pressure scale heights, much of the energy originally stored as the
enthalpy of the relativistic bubble is transferred to the gas.
Subsonic motion with respect to the ambient gas guarantees that only a
fraction of this energy is ``lost'' as sound waves, which may leave
the cluster core. Thus we can conclude that a fraction of energy of
order unity is transferred to the ICM.  This leads to the conjecture
that essentially all of the mechanical energy is dissipated in the
cluster core, which acts as a calorimeter of AGN activity
\citep{2002MNRAS.332..729C}.

Details of the dissipation process depend sensitively on the
properties of the ICM. For instance, this energy could drive turbulent
motions in the wake of the rising bubble and excite gravity waves
\citep[e.g.,][]{2001ApJ...554..261C,2004MNRAS.348.1105O}. These
motions will eventually dissipate into heat. Alternatively (if the ICM
viscosity is high), the energy can be dissipated directly in the flow
around the bubble. However in either case, the energy does not escape
from the cluster core.

\begin{figure}[t]
\includegraphics[width=\columnwidth]{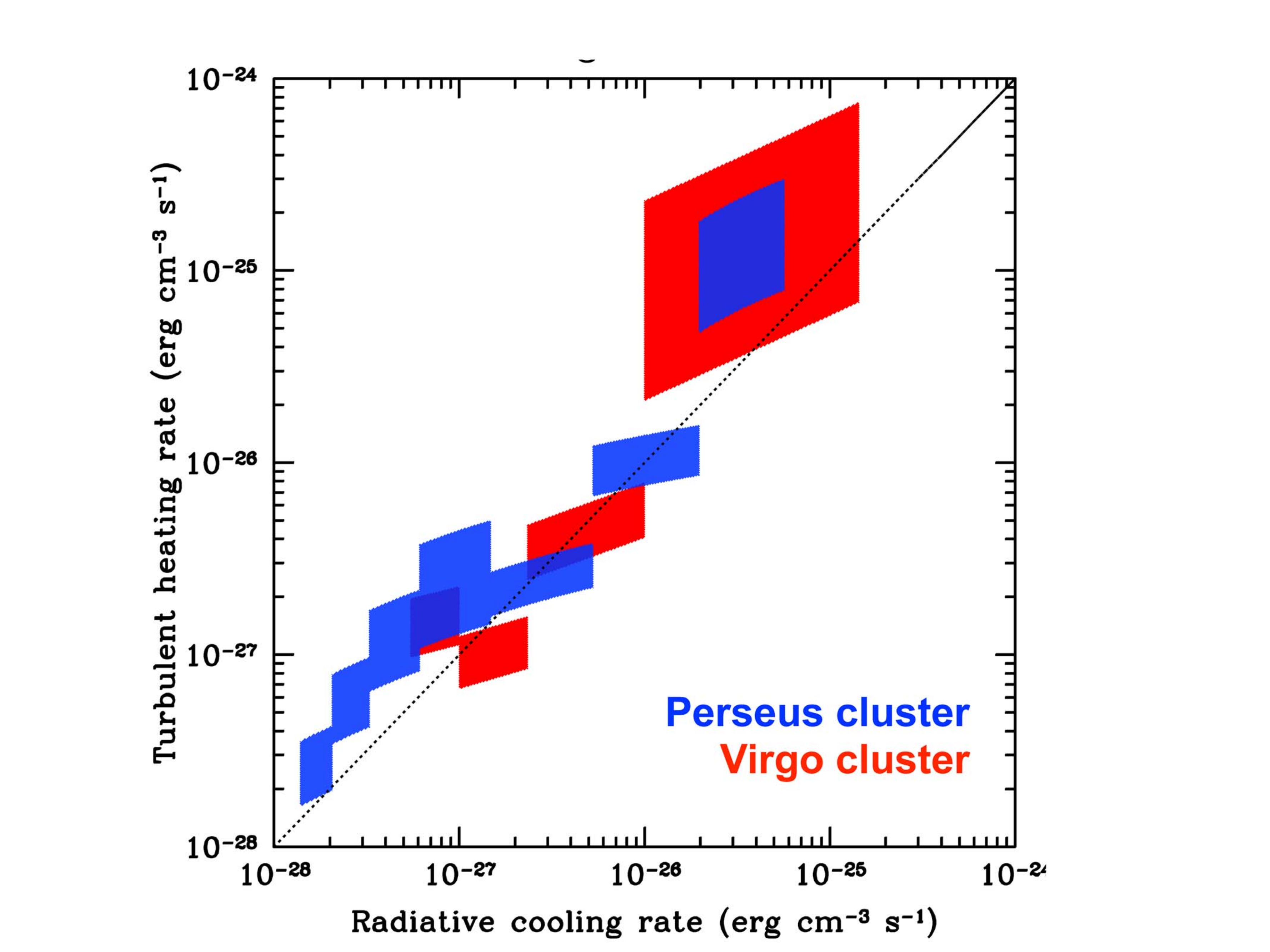}
\caption{Turbulent heating versus gas cooling rates in the
Perseus and Virgo cores. Shaded regions show the heating and cooling
rates estimated at different distances from the cluster center. The
size of each region reflects estimated statistical and stochastic
uncertainties.
\citep[Adapted from][]{2014Natur.515...85Z}.}
\label{fig:heating}
\end{figure}

An interesting recent development came from the analysis of X-ray
surface brightness fluctuations in the Perseus and M87/Virgo clusters
\citep{2014Natur.515...85Z}. If the observed fluctuations are
interpreted as weak perturbations of a nearly hydrostatic cluster
atmosphere, then one can link their amplitude to the characteristic
gas velocities at different spatial scales. Since in the canonical
Kolmogorov turbulence the energy flow is constant across the inertial
range, it is sufficient to know the velocity $V$ at one scale $l$
(within the inertial range) to estimate the dissipation rate as
$\displaystyle Q_{turb}\sim \rho \frac{V^3}{l}$. Applying this
approach to Perseus and M87/Virgo leads to a tantalizing conclusion
that the turbulent dissipation approximately matches the gas cooling
rate in these clusters (Fig.\ref{fig:heating}). While a number of
assumptions enter these calculations, the result is
encouraging. Future ASTRO-H measurements of the gas velocities in
these clusters should be able to verify these findings.

A substantial fraction of energy released by the SMBH can go into
quasi-spherical sound waves propagating through the ICM. Unlike strong
shocks the dissipation of the energy carried by sound waves depends on
the ICM microphysics, but it is plausible that their energy will be
dissipated before the wave leaves the core of the cluster
\citep[][]{2006MNRAS.366..417F}.  The attractiveness of this model is
that the energy can be evenly distributed over large regions. Whether
sound waves provide the dominant source of heat to the ICM depends
critically on what fraction of AGN energy goes to sound waves.  In
M\,87 the fraction of energy, which went into the weak shock is
$\lesssim$25\% \citep{2007ApJ...665.1057F}.

\subsection{Self-regulation}

\begin{figure*}
\includegraphics[width=0.99\columnwidth]{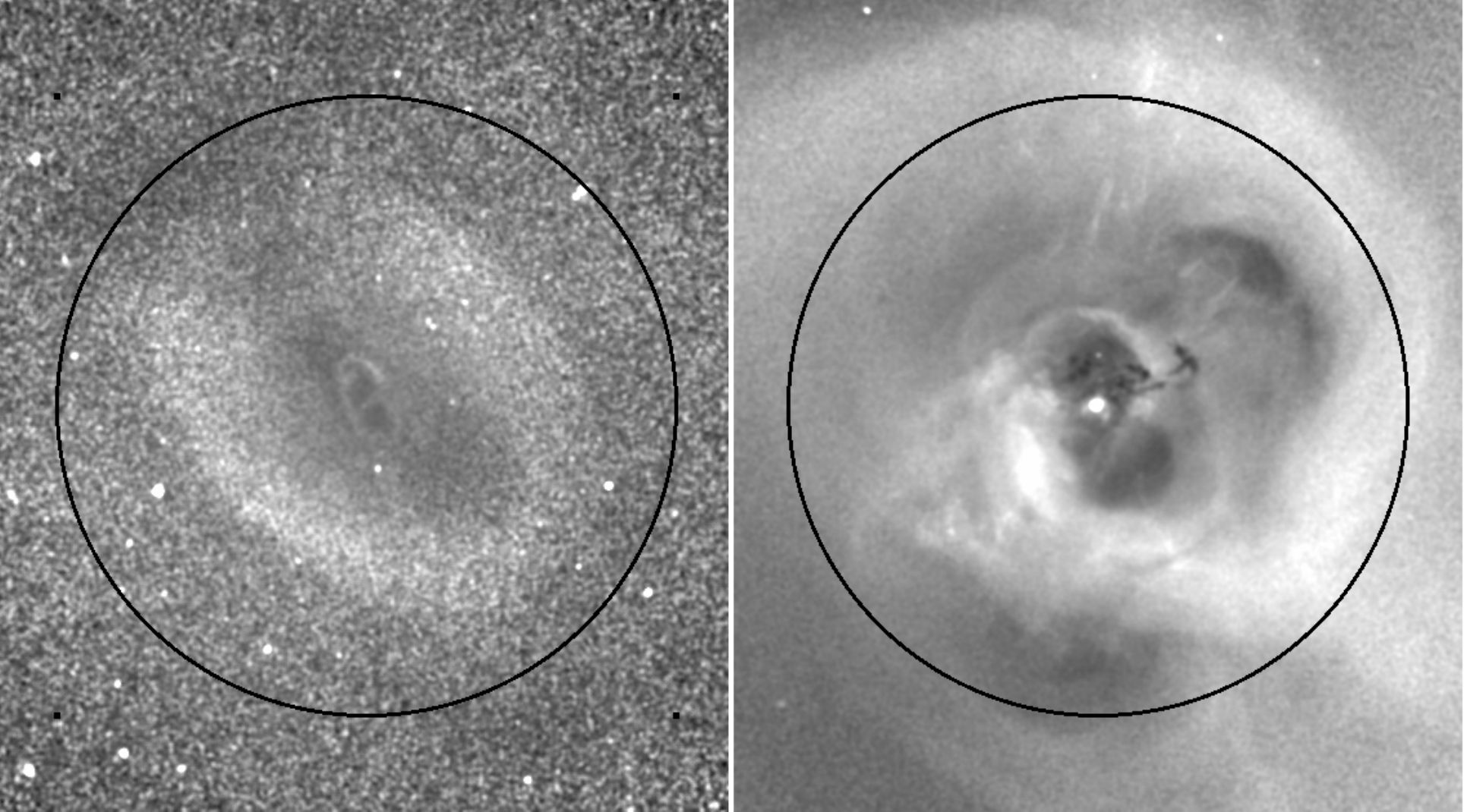}
\caption{Central $3' \times 3'$ region of the Chandra 0.6--2 keV
  images of NGC~5813 (left) and the Perseus cluster (right). Cavities
  inflated by AGNs are clearly visible in both images. For NGC~5813
  two (or even three) generations of cavities are easily identifiable. The volume of
  the cavities in Perseus is a factor of $\sim$10 larger than in NGC~5813.
\label{fig:2bub}}
\end{figure*}

Similar signs of SMBH-ICM interaction are observed for objects having
vastly different sizes and luminosities. Two examples are shown in
Fig.~\ref{fig:2bub} - these are $3' \times 3'$ X-ray images of the
elliptical galaxy NGC~5813 \citep{2011ApJ...726...86R} and the Perseus
cluster \citep{2000MNRAS.318L..65F}.  In each case, X-ray cavities of
approximately the same angular size are clearly seen. The distances to
NGC~5813 and the Perseus cluster are 32 and 70 Mpc,
respectively. Therefore the volumes of the cavities differ by a factor
of 10. A more extreme example is the MS~0735.6+7421 cluster
\citep{2005Natur.433...45M} at redshift $z$=0.22, which has cavities
with volume $\sim 10^4$ times larger than in NGC~5813.

A systematic comparison of the AGN mechanical power and the gas
cooling losses has been done for several dozens of objects
\citep[e.g.,][]{2006ApJ...652..216R,2012MNRAS.421.1360H}. These
studies suggest an approximate balance between AGN energy input and
cooling losses, implying that some mechanism is regulating the AGN
power to maintain this balance. A natural way to establish such
regulation is to link the accretion rate onto SMBHs with the
thermodynamic parameters of the gas. Two scenarios are outlined below.

In the first scenario, known as ``hot accretion'', the classic Bondi
formula \citep{1952MNRAS.112..195B} regulates the accretion rate of
the hot gas onto the black hole and provides a link between the gas
parameters and the mechanical power of an AGN
\citep[e.g.,][]{2002MNRAS.332..729C,2002A&A...382..804B,2003ApJ...582..133D}.

The rate of spherically symmetric adiabatic gas accretion without
angular momentum onto a point mass can be written as:

\begin{eqnarray}
 \dot{M}=4\pi \lambda (GM)^2 c_{\rm s}^{-3}\rho \propto s^{-3/2},
\label{bondi}
\end{eqnarray}

\noindent where { $\lambda$} is a numerical coefficient, which depends
on the gas adiabatic index $\gamma$ (for { $\gamma=5/3$} the maximal
valid value of $\lambda$ allowing steady spherically symmetric
solution is $\lambda_c=0.25$), $G$ is the gravitational constant, $M$
is the mass of the black hole, $c_{\rm s}=\sqrt{\gamma\frac{kT}{\mu
    m_{\rm p}}}$ is the gas sound speed, $\rho$ is the mass density of
the gas and $\displaystyle s=\frac{T}{n^{2/3}}$ is the entropy index
of the gas.

Thus the Bondi accretion rate is proportional to $ s^{-3/2}$ which
increases when the gas entropy decreases. Assuming that the heating is
directly proportional to the SMBH accretion rate (Heating$=\epsilon
c^2 \dot{M}$), there must be a value of gas entropy such, that heating
balances cooling:

\begin{eqnarray}
 s \approx 3.5 \left (\frac{M}{10^9  M_\odot} \right
)^{4/3} \left (\frac{\epsilon}{0.1} \right
)^{2/3}   \left ( \frac{L_X}{10^{43}~{\rm erg}\,{\rm s}^{-1} }
\right )^{-2/3} ~ {\rm keV}~{\rm cm}^2
\label{entropy}
\end{eqnarray}

\noindent A lower/higher entropy than this value implies too
much/little heating and therefore an overall increase/decrease of the
accretion rate. Since in stable hydrostatic equilibrium the low
entropy gas falls to the bottom of the potential well (the location of
the SMBH), the energy input is controlled by the minimum value of the
gas entropy in the whole central region.  This provides a natural
mechanism for self-regulation of the cooling and
heating. Interestingly, in many nearby systems this simple
prescription leads to an order of magnitude balance between cooling
and heating
\citep[e.g.,][]{2002MNRAS.332..729C,2002A&A...382..804B,2006MNRAS.372...21A}.

\begin{figure}
\includegraphics[width=0.99\columnwidth]{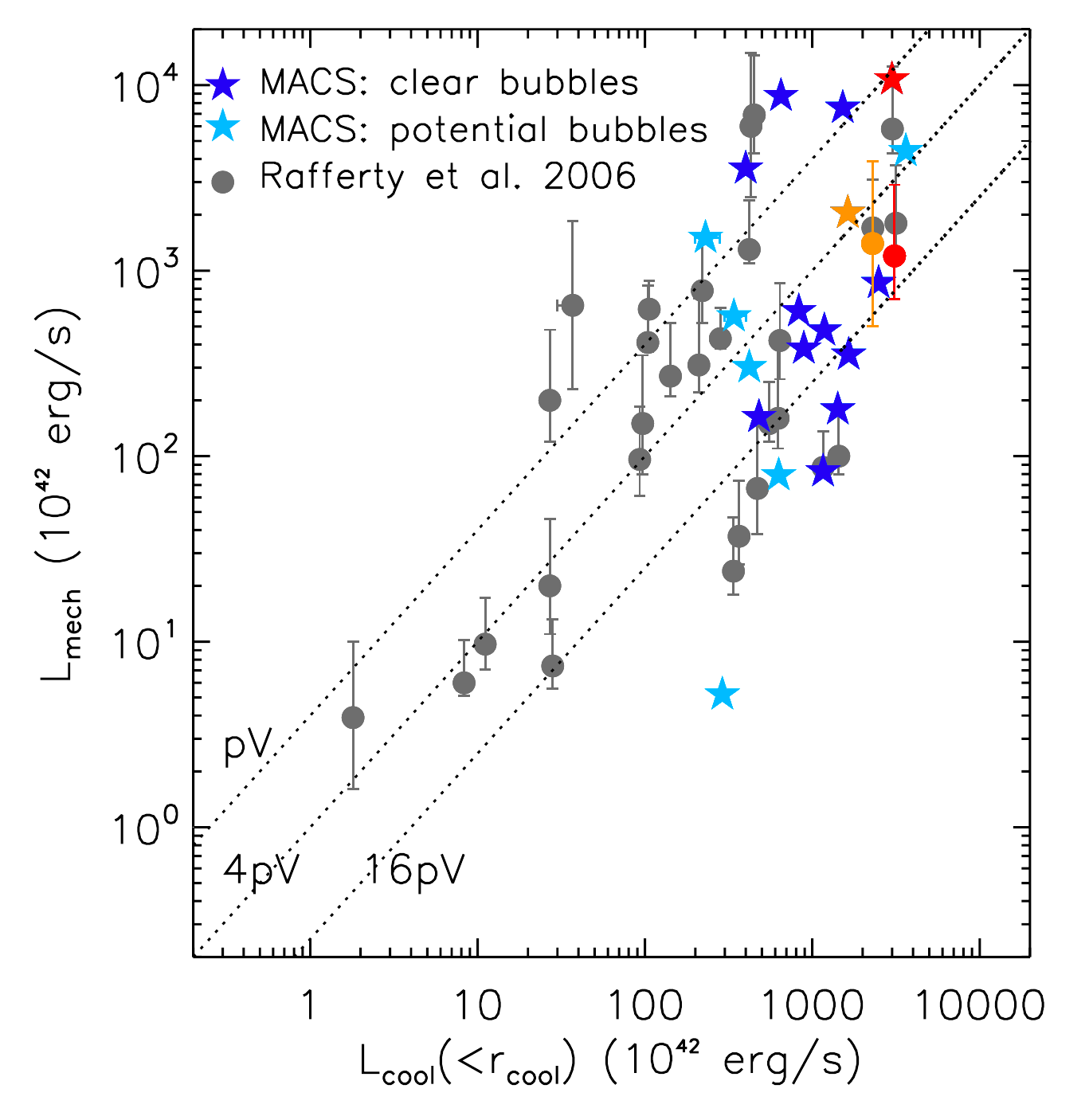}
\caption{Comparison of the estimated AGN mechanical power and the ICM
  cooling losses for a sample of clusters. Adopted from
  \citet{2012MNRAS.421.1360H}. The correlation is evident, albeit with
  substantial scatter, suggesting that mechanical output is regulated
  to match the cooling losses.
\label{fig:julia}}
\end{figure}

Another possibility is that some (sufficiently small) amount of gas is
first able to cool from the hot phase down to low temperatures
(e.g. down to $10^4$ K or below). Cold gas blobs then move in the
potential well of the central galaxy, collide, lose angular momentum
and feed the black hole
\citep[e.g.,][]{2005ApJ...632..821P,2012ApJ...746...94G}. This model
is known as a ``cold accretion'' scenario, as opposed to Bondi-type
accretion straight from the ``hot'' phase. While these scenarios
differ strongly in the physical process involved, they both advocate a
negative feedback loop, when the SMBH affects the thermal state of the
gas, which in turn affects the accretion onto the black hole.

\subsection{Link to evolution of galaxies}

A correlation of galaxy bulge properties and the mass of the SMBH
\citep[e.g.,][]{2000ApJ...539L...9F,2000ApJ...539L..13G} implies that
the black hole and its parent galaxy affect each other. The clear
evidence of mechanical feedback in nearby galaxy clusters suggests
that the same mechanism may be relevant for the formation of massive
ellipticals and for the growth of their SMBHs at $z\sim 2-3$.  Three
pre-requisites are needed for this scenario to work (i) a hot gaseous
atmosphere in the galaxy is present, (ii) the black hole is
sufficiently massive and (iii) large fraction of AGN energy is in
mechanical form and the coupling of mechanical energy to the ICM is
strong.

\begin{figure*}
\includegraphics[width= 0.75\textwidth]{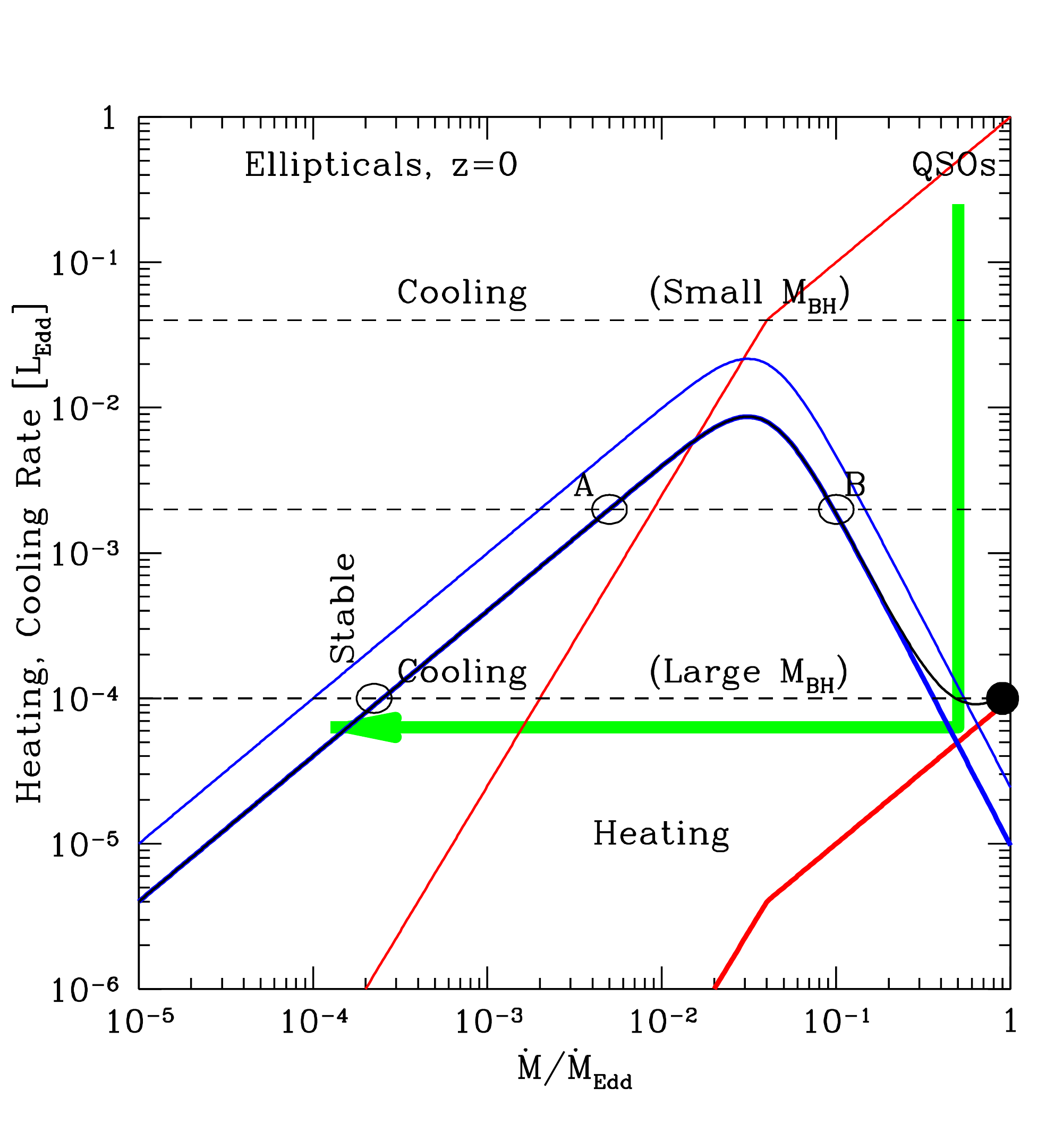}
\includegraphics[width=0.75\textwidth]{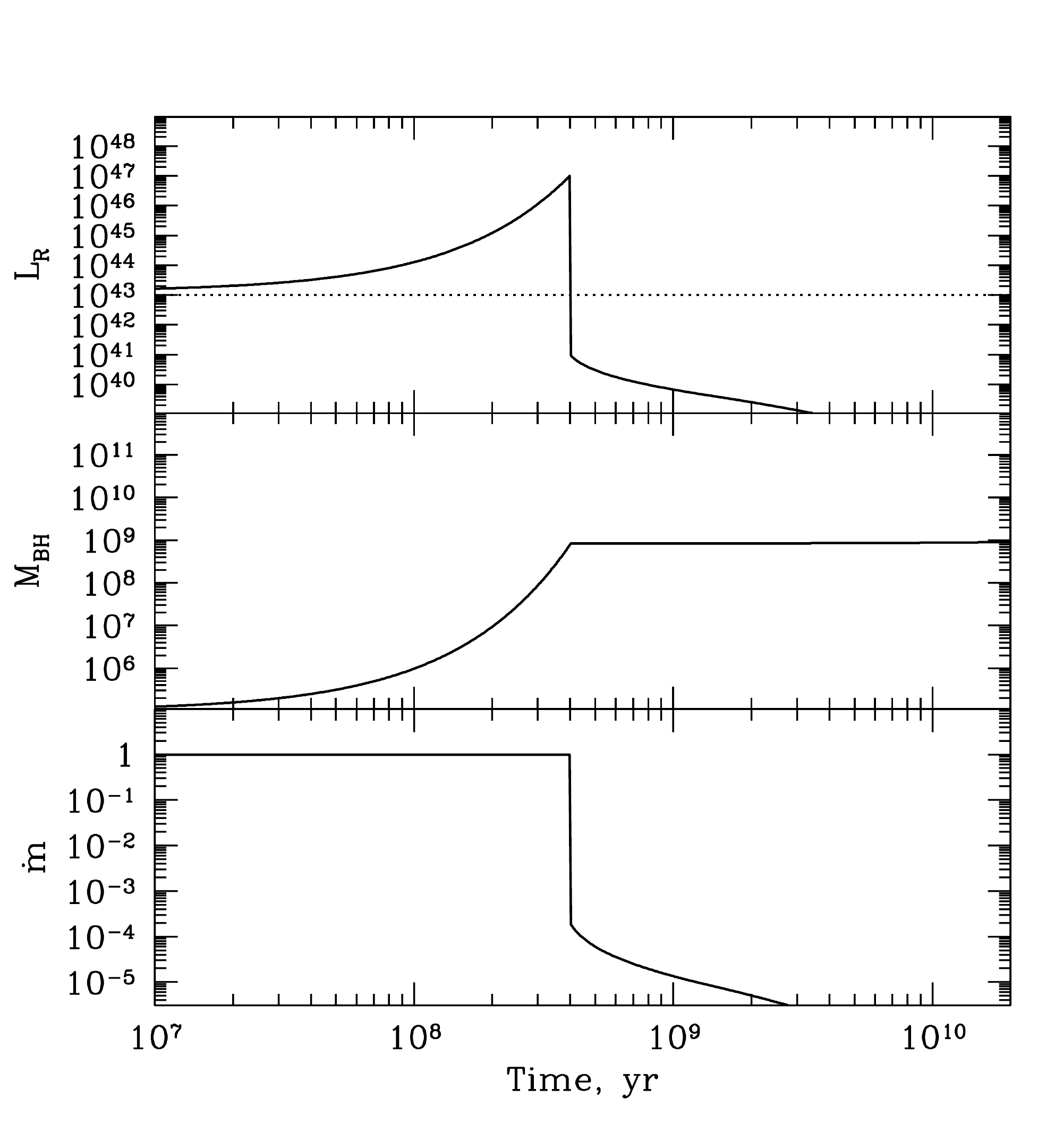}
\caption{{\bf Top:} Illustration of gas heating and cooling in elliptical
galaxies \citep{2005MNRAS.363L..91C}. The thick solid line shows, as a function
of the SMBH accretion rate, the heating rate due to outflow, which is
complemented/dominated by radiative heating near the Eddington limit. Horizontal
dashed lines show the gas cooling rate. The upper cooling line represents a
young galaxy in which a large amount of gas is present and/or the black hole is
small. Feedback from the black hole is not able to compensate for gas cooling
losses and the black hole is in the QSO stage with a near-critical accretion
rate, high radiative efficiency and weak feedback. As the black hole grows it
moves down in this plot. The black solid dot marks the termination of this
stage, when the black hole is first able to offset gas cooling, despite the low
gas heating efficiency. The lower cooling line illustrates present day
ellipticals: a stable solution exists at low accretion rates when mechanical
feedback from the black hole compensates gas cooling losses. The radiative
efficiency of accretion is very low and the black hole growth rate is very slow.
{\bf Bottom:} Possible time evolution, corresponding to the figure above: the
black hole accretes at the Eddington rate until its mass is large enough so that
even weak feedback does not allow a stable solution. At later times the AGN
switches into the low accretion rate mode, and the radiative efficiency drops
dramatically. }
\label{fig:accretion}
\end{figure*}

One can parametrize the magnitude of the feedback/gas heating $H$ with a simple
expression:

\begin{eqnarray}
H(M_{BH},\dot{M})=\left [  \alpha_M\epsilon_M(\dot{m}) +
  \alpha_R\epsilon_R(\dot{m}) \right ] 0.1 \dot{M}c^2,
\end{eqnarray}

\noindent where $\epsilon_M(\dot{m})$ and $\epsilon_R(\dot{m})$
characterize the efficiency of the transformation of accreted rest
mass per unit time $\dot{M}c^2$ into mechanical energy and radiation
respectively, while $\alpha_M$ and $\alpha_R$ are AGN-ICM coupling
constants -- the fraction of the released energy which is eventually
transferred to the gas. The first pair of coefficients
$(\epsilon_M(\dot{m}),\epsilon_R(\dot{m}))$ should come from accretion
physics, while the second pair $(\alpha_M,\alpha_R)$ depends on the
properties of the ICM and on the details of the AGN-ICM
interaction. The value of $\alpha_R$ is typically very low
$\alpha_R\lesssim 10^{-4}$
\citep{2004MNRAS.347..144S,2005MNRAS.358..168S}, while $\alpha_M$ can
be close to unity (see above). This difference between $\alpha_R$ and
$\alpha_M$ is the most important element of the mechanical feedback
scenario.

Let us assume that the system (SMBH + gaseous atmosphere) evolves to
the state where the heating by the black hole is equal to the gas
cooling losses (if such a state does exist). Thus the mass accretion
rate is the solution of the equation:

\begin{equation}
H(M_{\rm BH},\dot{M})=L_{\rm cool},
\label{eq:hec}
\end{equation}

\noindent provided $\displaystyle \dot{m}=\frac{\dot{M}}{\dot{M}_{\rm
    Edd}}\le 1$. In the opposite case, we set $\dot{M}=\dot{M}_{\rm
  Edd}$. In other words, if the black hole cannot offset ISM cooling
losses even at the Eddington rate then it will keep accreting gas at
this rate. If there is a solution of Eq.\ref{eq:hec} at $\dot{m}\le 1$
then at this rate an equilibrium between heating and cooling is
possible. We will see below that several distinct solutions of
Eq.~\ref{eq:hec} are possible (some of them are unstable).

Let us now assume that the mechanical output $\epsilon_M(\dot{m})$ is
large at low accretion rate, but decreases at high accretion rates
(so-called ``radio mode''). The radiative output $\epsilon_R(\dot{m})$
is on the contrary large at high accretion rates and decreases at low
accretion rates. The mechanical and radiative outputs corresponding to
this scenario are shown in Fig.  \ref{fig:accretion} with the thin
blue and red curves respectively.  The above assumptions are motivated
by X-ray and radio observations of several X-ray binaries
\citep[e.g.,][]{2003MNRAS.344...60G} and AGNs
\citep[e.g.,][]{2000ApJ...543..611O} and can also be supported by
theoretical arguments for radiatively inefficient accretion flows
\citep[e.g.,][]{1977ApJ...214..840I,1982Natur.295...17R,1994ApJ...428L..13N}.
For instance, the nucleus of M87 can be regarded as the prototypical
example of an AGN in the low accretion rate mode, when kinetic power
of radio emitting outflows exceeds its radiative power. At the same
time, there are black holes in binary systems in the high accretion
regime, which are very bright in X-rays, but show no evidence for a
powerful outflow.

We now set $\displaystyle \alpha_M \sim 0.7$ and $\displaystyle
\alpha_R \sim 10^{-4}$. The ICM heating by mechanical and radiative
power (obtained by multiplying the curves by 0.7 and $10^{-4}$) is
shown by the thick blue and red curves respectively. Finally the total
ICM heating rate (the sum of two curves) is plotted as a thick black
curve. Once the black hole is sufficiently massive, there are two
possible solutions with the heating balancing ICM cooling losses
(points marked as A and B in Fig. \ref{fig:accretion}). The point B is
certainly unstable since an increase of the mass accretion rate causes
a decrease of the heating rate. The system will evolve from state B
into one with lower or higher accretion rate. If the system goes into
a high accretion rate mode then the feedback power drops and enhanced
cooling boosts the accretion rate towards the Eddington
value. Finally, the black hole mass reaches the level where AGN
heating exceeds ICM cooling at any (sufficiently high) accretion
rate. The gas entropy increases, the mass accretion rate drops and the
system then switches to the stable state at low accretion rate, low
radiative efficiency and high mechanical efficiency. This is the so
called ``radio-mode'' which presumably describes M87 now. In the
simple scenario outlined above the black hole at the center of the
cooling core first looks like a QSO and then jumps to a state of a low
luminosity AGN as shown with green line in Fig. \ref{fig:accretion}.

The time evolution of the black hole mass, accretion rate and
radiative luminosity is sketched in the right panel of
Fig. \ref{fig:accretion}. The black hole initially accretes at the
Eddington rate and grows exponentially with limited impact on the
ICM. This fast growth and QSO-type behavior of the black hole
continues until its mass is large enough so that even with pure
radiative feedback, heating can offset cooling losses. After the
system moves to the low accretion state the luminosity and the black
hole growth rate drops by a large factor of order the ratio of the
coupling constants $\alpha_M/\alpha_R \sim 10^4$ (or more if an
ADAF-type scenario is adopted).

The above consideration is of course an overly simplified
picture. Various prescriptions of the ``radio-mode'' feedback have
been tested \citep[e.g.,][]{2006MNRAS.365...11C,2006MNRAS.370..645B}
in the semianalytic models coupled with the numerical simulations of
structure formation of the Universe. In general, the key element of
mechanical AGN feedback -- the ability to provide good coupling of the
AGN and the ICM -- seems to be able to resolve the difficult issue of
over-cooling and excess star formation in the most massive halos.

\subsection{Conclusions about AGN feedback}

AGN Feedback in galaxy clusters is a rapidly developing area in
astrophysics. It depends on a combination of various physical
processes, ranging from physics of accretion to cosmological evolution
of the most massive systems in our Universe.  Much of the physics
involved is still poorly understood. At the present epoch, AGN
feedback prevents gas cooling in massive elliptical galaxies and
clusters, but its role at higher redshifts is only beginning to emerge
today \citep[e.g.][]{2012MNRAS.421.1360H,2013ApJ...774...23M}. All
this leaves much opportunity for future observational and theoretical
studies.

\section{Galactic feedback and metal enrichment in clusters}
\label{sec:3.1.2}

During the formation history of clusters of galaxies, metals have been
continuously produced by the stars in the member galaxies. In the
earliest epoch of star formation, about 500 Myr after the 'Big bang',
Population III stars formed from the primordial gas. The nature of
this stellar population is still uncertain, but probably it consisted
of intermediate and high-mass stars \citep{vangioni2011}. The first
metals produced by this population were mostly ejected into the
surrounding medium, leading to an initial metal abundance of about
10$^{-4}$ times Solar \citep{matteucci2005}. This relatively low metal
abundance was enough to allow the gas to cool more efficiently through
spectral line emission, resulting in an epoch of increased star
formation, which peaked around a redshift of $z \sim 2-3$.

This peak of the Universal star formation rate roughly coincides with
the build-up of a hot Intra-Cluster Medium in the massive galaxy
clusters. Dilute metal-poor gas accreted from the cluster surroundings
is mixed with the gas expelled by galactic winds driven by supernova
explosions. A combination of compression, accretion shocks, supernova
heating and AGN feedback boosted the ICM temperature to more than 1
million Kelvin. In such a hot environment, star formation in the
cluster was effectively quenched \citep{gabor2010}. While the build-up
of the ICM continued, the star formation rate dropped to very low
levels, which explains the old stellar populations observed in local
galaxy clusters today.

Due to their deep gravitational potential wells, clusters have
retained the metals ejected into their ICM. Metal abundances observed
in the ICM of local clusters thus provide an integrated record of
their enrichment history. The low star-formation rates in cluster
galaxies since $z \sim 2-3$ form an interesting contrast to the
enrichment history of our own galaxy, where the star-formation rate
showed a smaller decline. Clusters of galaxies therefore provide a
unique insight in the chemical enrichment due to stellar populations
before $z \sim 2$.

\subsection{Sources of metals}

\begin{figure}[t]
\includegraphics[width=0.7\columnwidth,angle=0]{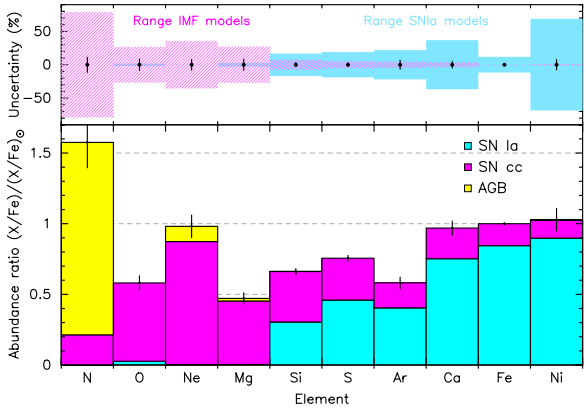}
\caption{Expected abundances measured in a 120 ks XMM-Newton observation of
S\'ersic 159-03 (bottom panel), which is a typical bright local cluster. The
statistical error bars were obtained from \citet{deplaa2006}. The estimates for
the SNIa, SNcc, and AGB contributions are based on a sample of 22  clusters
\citet{deplaa2007} and two elliptical galaxies \citep{grange2011}. The top
panels show the typical range in SNIa and IMF models with respect to the
statistical error bars in the observation.}
\label{fig:barplot}
\end{figure}

Most of the elements heavier than beryllium are produced in
supernovae. Some elements, like nitrogen and sodium, are ejected into
the medium by Asymptotic Giant Branch (AGB) stars. There are two
distinct types of supernova explosions, type Ia and core-collapse,
that each have a separate role in metal enrichment.  Core-collapse
supernovae yield the bulk of the elements in the mass range from
oxygen to silicon, while type Ia supernovae produce mostly elements
from silicon to nickel (see Fig.~\ref{fig:barplot}).

Supernova type Ia models, however, do not agree well on the exact
yields for each element. Uncertainties on the yields of, for example,
calcium and nickel are about 40--60\%. The underlying cause is the
very interesting progenitor problem of type Ia supernovae. Recent
optical observations of type Ia supernovae show variations in their
spectral properties. So far, it has been challenging to link an
observed type Ia supernova to a progenitor type \citep[see
e.g.][]{howell2011}. Two main progenitor scenarios are being
considered. The first is the 'classical' type Ia, where a white dwarf
accretes matter from a companion star in the Red Giant phase. If the
accretion rate is right, the mass and temperature of the white dwarf
can grow to a point where carbon fusion ignites. This is close to the
Chandrasekhar limit of 1.4 solar masses. The carbon ignites
explosively and unbinds the entire white dwarf. The second progenitor
channel is a scenario where two white dwarfs merge. They spiral toward
each other through the emission of gravitational radiation and the
less massive star is accreted onto the more massive one, until carbon
is ignited and the star explodes. It is clear that these progenitor
channels allow a range of possible explosion scenarios and metal
yields, which makes accurate predictions challenging.

The yields of core-collapse supernovae are currently better
established. The main uncertainty in the core-collapse yield of an
entire stellar population is the Initial-Mass Function (IMF). To
obtain the total yield for a population, the yields calculated for
individual masses need to be integrated over the IMF
\citep{tsujimoto1995}. Therefore, the observed abundances of metals in
the oxygen to silicon mass range can constrain the IMF of the
cluster's stellar population above $\sim$8 solar masses.

Below 8 $M_{\odot}$, intermediate-mass stars in their AGB phase are a
source of nitrogen and sodium. Yields for these sources for several
mass bins are available \citep{karakas2010} and also need to be
integrated over the IMF, like core-collapse yields, to obtain the
yield for the whole stellar population.  Measuring the abundances of
AGB products therefore puts constraints the low-mass end of the IMF.

\subsection{Abundance measurements in X-rays}

The soft X-ray band between 0.1 and 10 keV is very suitable for
abundance studies because it in principle contains spectral lines from
all elements between carbon and zinc. In clusters of galaxies, the hot
plasma is in (or is very close to) collisional ionisation equilibrium,
which makes it relatively easy to predict the emitted X-ray
spectrum. The advance of X-ray spectroscopy has enabled the study of
abundances in clusters and yielded interesting results \citep[See
e.g.][for more extended reviews on this topic]{werner2008,deplaa2013}.

The first attempt to link cluster abundances to supernova yields was
done with the ASCA satellite \citep{mushotzky1996}. With this
instrument it was already possible to measure the abundances of O, Ne,
Mg, Si, S, Ar, Ca, Fe and Ni. From these, and other ASCA studies the
picture emerged of an ICM that was enriched early in its formation
history with core-collapse supernova products and later by type Ia
supernovae.

\begin{figure}[t]
\includegraphics[width=\columnwidth]{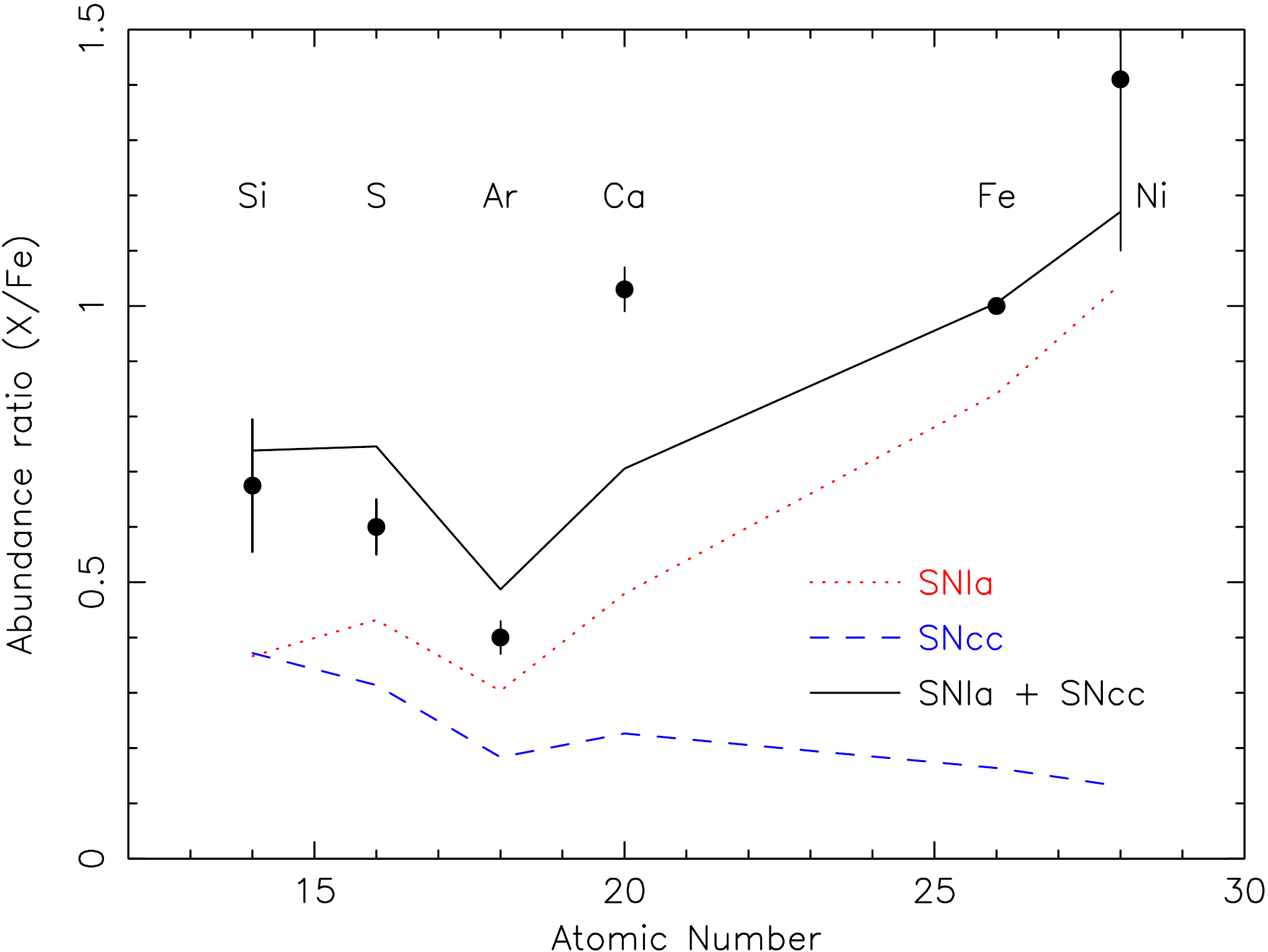}
\caption{Abundance ratios fitted with supernova yields from the WDD2 SNIa model
\citep{iwamoto1999} and a SNcc model with an initial metallicity $Z=0.02$ and a
Salpeter IMF. The calcium abundance appears to be underestimated. It can not
explain the Ar/Ca ratio measured in this XMM-Newton sample of 22 clusters.
\citep[Adapted from][]{deplaa2007}.}
\label{fig:calcium}
\end{figure}

With the launch of XMM-Newton, a telescope with a substantial
effective area and spectral resolution (through the Reflection Grating
Spectrometer, RGS) became available, which enabled deep abundance
studies in larger cluster samples. In de \citet{deplaa2007}, for
example, 22 clusters were analysed and abundances were measured in
their core regions. If the average abundances of the sample are
compared to supernova type Ia models, the calcium abundance appeared
to be systematically higher than expected by the models (see
Fig.~\ref{fig:calcium}). Possible explanations for this high calcium
abundance include an unexpected difference in the type Ia explosion
mechanism or an increased importance of progenitor systems where
helium is accreted on the white dwarf, because explosive He fusion is
expected to yield more calcium. Although a systematic error in the
determination of the Ca abundance cannot yet be fully excluded
(detailed analysis did not show any problem), this measurement can
constrain supernova type Ia models. Note that the fit shown in Fig.~\ref{fig:calcium}
includes Ca. If Ca is excluded  a better fit for S and Ar is obtained.

If a combination of a type Ia model and core-collapse model fits,
their ratio is an indication of the relative contribution of type Ia
supernovae to the enrichment. \citet{bulbul2012} developed an
extention to the APEC model that is able to fit the type Ia to
core-collapse ratio directly to the spectra. This ratio depends
strongly on the used models \citep{degrandi2009}. But despite of this
uncertainty, multiple groups \citep[e.g.][]{bulbul2012,sato2007}
report a type Ia contribution of $\sim$30-40\%, which is consistent
with optical data.

In cool clusters and groups, the RGS spectrometer aboard XMM-Newton is
able to measure carbon and nitrogen abundances. These elements are not
produced in large quatities in supernovae, but appear to originate
from metal-poor massive stars or AGB stars. The exact origin is still
subject of debate \citep{romano2010}.  RGS observations by
\citet{werner2006b} and \citet{grange2011} have shown nitrogen to be
very abundant around elliptical galaxies. The high abundance of
nitrogen cannot be explained by supernovae alone. A population of
intermediate-mass AGB stars is therefore likely responsible for the
high nitrogen content of the ICM.

With the Japanese Suzaku satellite, launched into low-earth orbit in
2005, abundance studies are performed in the outskirts of clusters
 \citep[see][for a recent and complete review of
  Suzaku interesting results]{2013SSRv..177..195R}.  In these low
surface-brightness areas, the low Suzaku background level is
favourable above XMM-Newton. Recently, in an elaborate study of the
outskirts of the Perseus cluster with Suzaku, \citet{werner2013} found
that the iron abundance distribution is surprisingly smooth in the
outskirts. This points toward a scenario of an early enrichment of the
hot X-ray gas with iron, well before $z \sim 2$. This result confirmes
earlier indications that chemical enrichment occured very early in the
development of the universe and probably also before the formation of
clusters.

\begin{figure}[t]
\begin{center}
\includegraphics[width=0.8\textwidth]{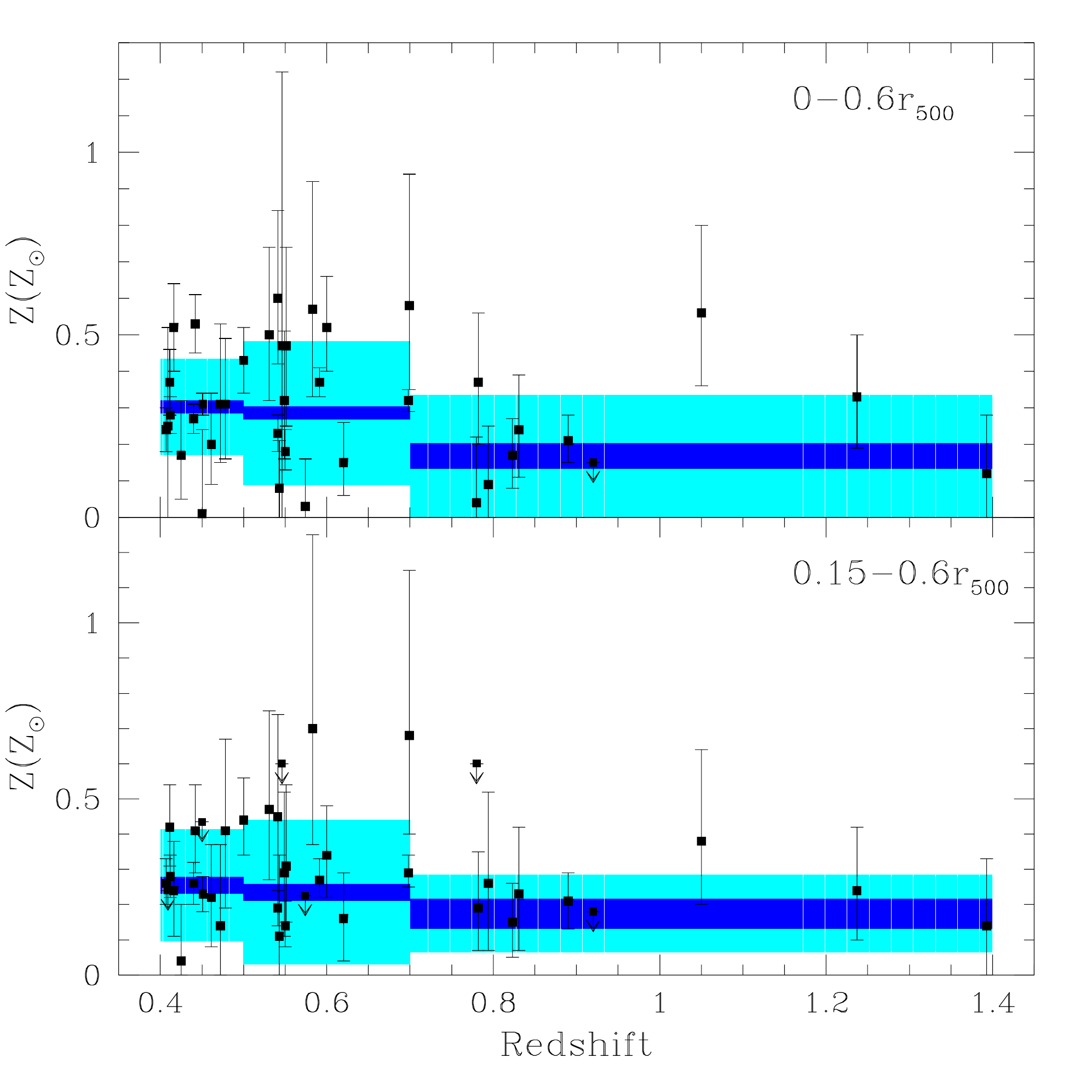}
\end{center}
\caption{Measured cluster abundances versus redshift. The top panel shows the
abundances up to 0.6r$_{500}$ and the lower panel shows the abundances when the
inner region up to 0.15r$_{500}$ is ignored. Adapted from \citet{baldi2012}.}
\label{fig:baldi}
\end{figure}

One of these earlier indications was found in a study of a sample of
high redshift clusters between $z=0.3-1.3$ by
\citet{baldi2012}. Although the scatter in the measured metal
abundance as a function of redshift is large, a significant trend in
the metal abundance up to $z=1.3$ was not found
(Fig.~\ref{fig:baldi}), indicating that the chemical enrichment
mechanisms have not added a lot to the enrichment since
$z=1.3$. Deeper observations of high redshift clusters would be
necessary to build a large enough sample to confirm the lack of a
metallicity trend with redshift.

\section{Magnetic feedback processes}

Galaxies play a key role in the enrichment of the ICM or IGM, not only
as far as heavy elements are concerned, but also regarding the
magnetization. According to the standard bottom-up scenario of lowest
galactic masses, primeval galaxies must have injected much of their
ISM into the IGM during the initial bursts of star formation, thereby
''polluting'' large volumes of intergalactic space because of their
high number density.

\citet{1999ApJ...511...56K} were the first to raise the question
whether low-mass galaxies could have made a significant contribution
to the magnetization of the IGM (apart from more massive starburst
galaxies and AGN).  Owing to their large number (observed and
predicted in a CDM cosmology) and their injection of relativistic
particles, they could have played a cardinal role in the context of
this cosmologically important scenario.  If true, it is to be expected
that dwarf galaxies are ``wrapped'' in large envelopes of previously
highly relativistic particles and magnetic fields, which are pushed
out of them during epochs of vigorous star formation.
\citet{2006MNRAS.370..319B} have discussed this more quantitatively
and made predictions for the strengths of magnetic seed fields to be
then amplified by large-scale dynamos over cosmic time. In particular,
they also predict the existence of magnetic voids.

\citet{2009MNRAS.392.1008D} and \citet{2010A&A...523A..72D} performed
numerical models of supernova-driven winds in dwarf galaxies. Their
simulations can provide an understanding of the origin of
intergalactic magnetic fields at the level of
$10^{-4}~\mu$G. \citet{2013MNRAS.435.3575B} presented a first
numerical model of supernova-driven seeding of magnetic fields by
protogalaxies.

The existence of winds in low-mass galaxies is inferred from the
observed kinematics of the gas (measured with slit spectroscopy), but
can arguably be also inferred from measurements of the temperature of
the hot (X-ray-emitting) gas. For instance,
\citet{1998ApJ...506..222M} found the outflow velocities in NGC\,1569
to exceed the escape speed, and \citet{1996ApJ...469..662D} derived
the temperature of its hot, X-ray-emitting gas to exceed the virial
temperature. The transport of a relativistic plasma out of this galaxy
is strongly suggested by radio continuum observations of dwarf
galaxies. \citet{2010ApJ...712..536K} studied the radio halo in
NGC\,1569, which extends out to about 2~kpc at 1.4~GHz. The dwarf
irregular NGC\,4449 also possesses a low-frequency radio halo
\citep{1996A&A...313..396K}.

Of course, in magnetizing the ICM/IGM, low-mass galaxies have been
competing with AGN \citep{1987QJRAS..28..197R}. Judging from the radio
luminosities of the ``culprits'', it is clear that nevertheless
low-mass galaxies may have contributed significantly. While a typical
starburst dwarf galaxy emits a monochromatic radio luminosity of
$P_{\rm 1.4~GHz} \approx 10^{20.5}$~W~Hz$^{-1}$, radio galaxies in the
FRI/II transition regime have luminosities of $P_{\rm 1.4~GHz} \approx
10^{24.7}$~W~Hz$^{-1}$.  Hence, the radio power produced by AGN is
about$10^4$ times larger than that of dwarf galaxies. On the other
hand, $\Lambda$CDM cosmology with bottom-up structure formation
implies that dwarf galaxies must have been formed in huge numbers.

The role of massive black holes in the build-up of strong magnetic
fields in galaxies was addressed by \citet{1994Natur.368..434C}. They
pointed out that galactic winds or collimated jets were able to
disperse such magnetic fields over large volumes of the host galaxies
and beyond. \citet{2010ApJ...725.2152X} presented
magneto-hydrodynamical simulations in which they studied the evolution
of magnetic fields ejected by an AGN shortly after the formation of a
galaxy cluster. They showed that, if the magnetic fields are ejected
before any major mergers occurring in the forming cluster, they can be
spread throughout the cluster, with further subsequent amplification
by turbulence in the ICM. It should be noted at this point that
central so-called ``mini-halos'', which are bright extended radio
sources located in the centers of cooling-flow clusters
(e.g. Perseus\,A, Hydra\,A, Virgo\,A), cannot possibly magnetize large
cluster volumes, as they are pressure-confined
\citep[e.g.][]{2012A&A...547A..56D}.


\section{Cold fronts in galaxy clusters}

\begin{figure*}[t]
\centering
\includegraphics[height=0.47\linewidth,bb=0 15 490 511,clip]%
{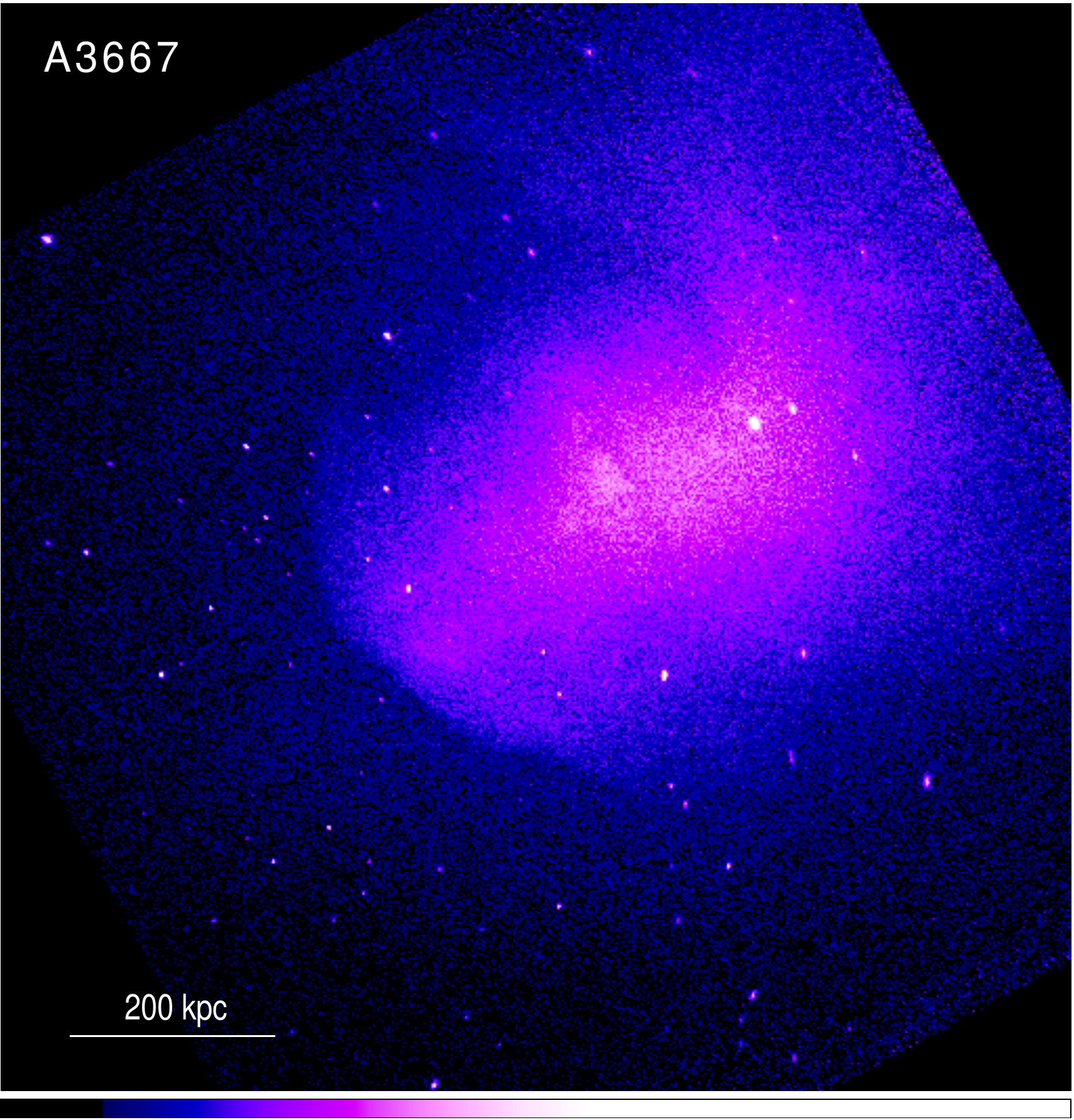}
\hspace{2mm}
\includegraphics[height=0.47\linewidth,bb=53 223 562 728,clip]%
{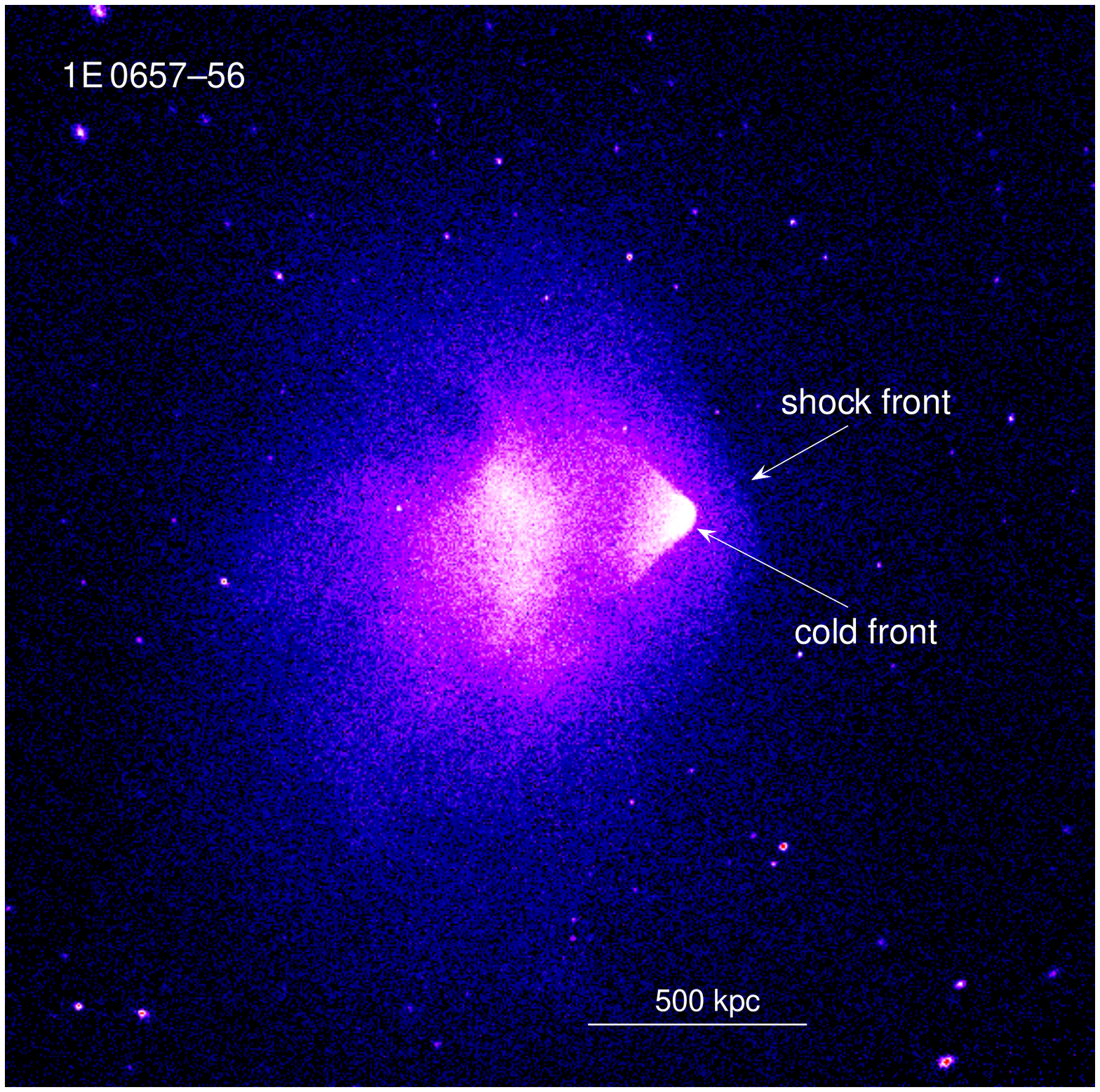}

\caption{Prominent cold fronts in the Chandra X-ray images of A\,3667 (MV07)
  and the Bullet cluster \citep{markevitch06}. In the Bullet cluster, the
  surface of the cool ``bullet'' is a cold front; its supersonic motion
  generates a prominent bow shock.}
\label{fig:a3667_1e}
\end{figure*}

\begin{figure*}[t]
\centering
\includegraphics[width=0.7\linewidth,bb=1 22 340 405,clip]%
{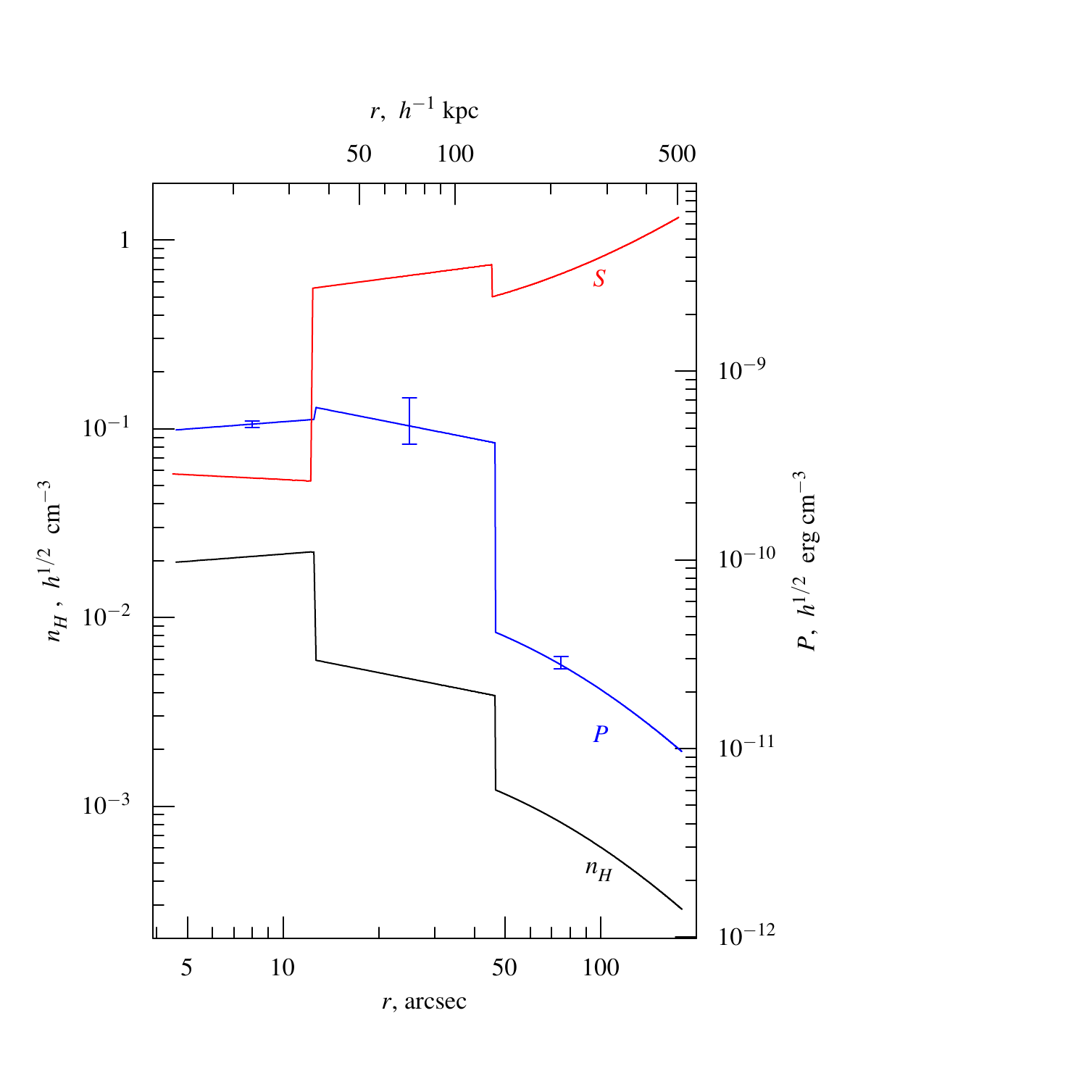}

\caption{Schematic radial profiles of the gas density, pressure and specific
  entropy in a sector crossing the bullet and the bow shock in the Bullet
  cluster and centered on the bullet's center of curvature (see Fig.\
  \ref{fig:a3667_1e}). The shock front and the bullet boundary have sharp
  density jumps of similar amplitudes; however, the shock at $r\approx 50''$
  exhibits a large pressure jump and a slight entropy increase, while the
  bullet boundary at $r\approx 12''$ is in near pressure equilibrium but
  separates gases with very different entropies --- it is a cold front.
  (reproduced from MV07)}
\label{fig:1e_profs}
\end{figure*}

Among the first results from high-resolution cluster images obtained
with Chandra was the discovery of sharp edges in the X-ray surface
brightness in merging clusters A\,2142 and A\,3667
\citep{markevitch00} (M00); \citep{V01} (V01).
Fig.~\ref{fig:a3667_1e} shows deep Chandra images of A\,3667 and the
Bullet cluster, both of which exhibit such brightness edges.  The
characteristic shape of the brightness profiles across these edges
corresponds to a projected abrupt jump of the gas density at the
boundary of a roughly spherical body (M00). The Bullet cluster shows
two prominent edges, one at the nose of the dense cool ``bullet'' and
another ahead of it.  Their physical nature is revealed by radial
profiles of the gas density, pressure and specific entropy, shown in
Fig.~\ref{fig:1e_profs}. As one can guess already from the image, the
outer edge is a bow shock --- the dense side of the edge is hotter,
with the pressure jump satisfying the Rankine-Hugoniot jump conditions
\citep{markevitch06}.

At the same time, the boundary of the bullet has a different physical
nature --- it separates two gas phases with very different specific
entropies that are in approximate pressure equilibrium at the
boundary. The denser side of this edge has a lower temperature,
opposite to what's expected for a shock front, which is how the two
phenomena can be distinguished observationally. The edge in A\,3667
and those in A\,2142 have the same sign of the temperature jump as
that at the bullet boundary. These features in clusters they have been
named ``cold fronts'' (V01). The term ``contact discontinuity'' is
sometimes used, but it implies continuous pressure and velocity
between the gas phases, whereas these structures in clusters may have
discontinuous tangential velocity, as we will see below. Cold fronts
turned out to be much more ubiquitous than shocks and have been
observed with Chandra and XMM-Newton in many clusters and even
galaxies (e.g., \citet{machacek05}; see \citet{markevitch07} (MV07),
for a detailed review). If the dense gas cloud is moving with respect
to the ambient gas, there will be a ram pressure component
contributing to the pressure balance near the front (M00), which makes
it possible to estimate the velocity of the cloud (V01).

\begin{figure*}[t]
\centering
\includegraphics[width=0.47\linewidth,bb=57 220 564 729,clip]%
{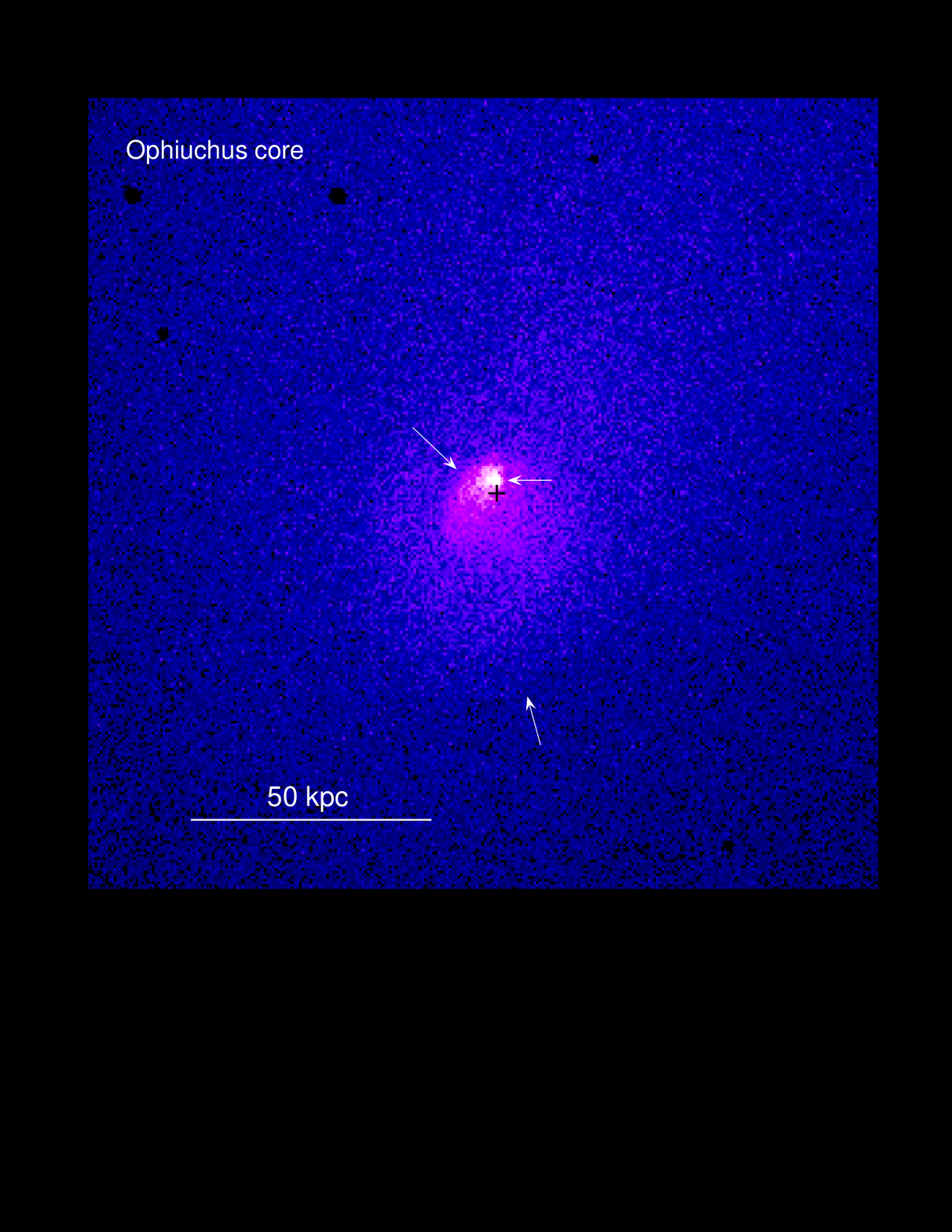}
\hspace{2mm}
\includegraphics[width=0.47\linewidth,bb=116 349 473 706,clip]%
{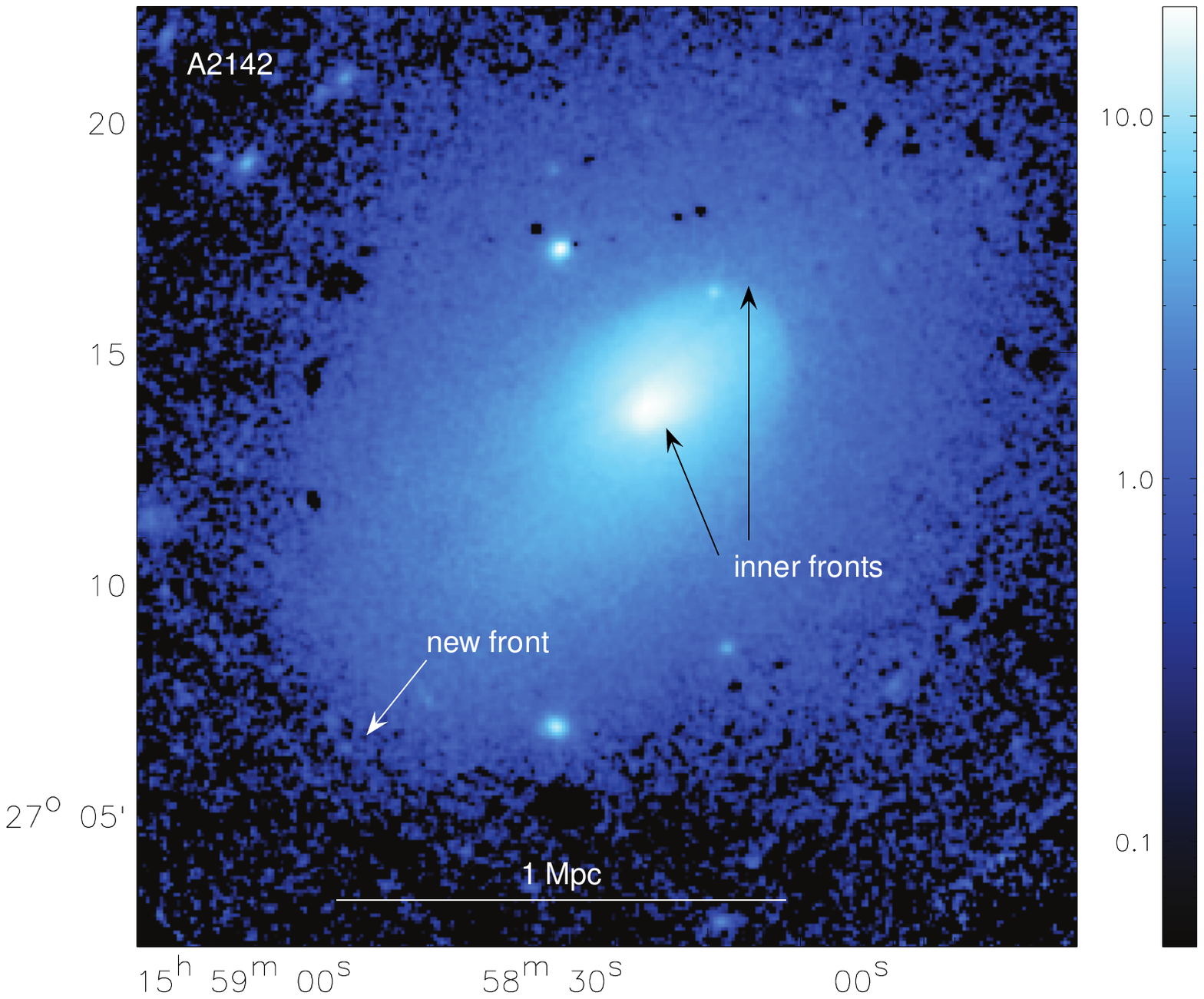}

\caption{{\em Left}: A series of ``sloshing'' cold fronts at different radii
  in the cool core of the Ophiuchus cluster, shown by arrows in this
  archival Chandra image. (The subtle outermost front is seen better in a
  more coarsely binned image.) Black cross marks the center of the cD
  galaxy.  {\em Right}: A cold front discovered in A\,2142 at a distance of 1
  Mpc from the center (marked ``new front'') by \citet{rossetti13} in
  this XMM-Newton dataset. The original cold fronts in the central region (M00)
  are shown by black arrows.}
\label{fig:oph_a2142}
\end{figure*}

Cluster cold fronts have two main physical causes, both related to
mergers.  The obvious one, originally proposed by M00 for the two
fronts in A\,2142, involves an infalling subcluster ram-pressure
stripped of its outer gas layers, which leaves a sharp boundary
between the dense remnant of the subcluster's core and the less dense,
hotter gas of the main cluster flowing around it.  The bullet in the
merging Bullet cluster and the NGC\,1404 galaxy falling into the
Fornax cluster \citep{machacek05} appear to be such ``stripping''
fronts.%
\footnote{It appears that the original M00 scenario for A2142, which
  involved two surviving gas cores, is not correct ( \citet{tittley05}; MV07).  Instead, the two original fronts, the front discovered at a
  large radius \citep{rossetti13}, and yet another front seen close to
  the center in a deep Chandra observation --- all concentric --- are
  ``sloshing'' fronts discussed below. The ``stripping'' scenario does work
  in some other clusters.}
However, cold fronts have also been observed in the cores of the majority of
{\em relaxed}\/ clusters that show no signs of recent merging
(\citet{mazzotta01}; \citet{markevitch01}, M01; \citet{markevitch03};
\citet{ghizzardi10}). These fronts are typically more subtle in terms of the
density jump than those in mergers, and occur close to the center ($r\lsim 100$
kpc), with their arcs usually curved around the central gas density peak. An
example is seen in the Ophiuchus cluster (Fig.\ \ref{fig:oph_a2142}a). A
detailed study of such a front in A\,1795 by M01 has shown that the gas on two
sides of the front has different centripetal acceleration. This led those
authors to propose that the dense gas of the cool core is ``sloshing'' around
the center of the cluster gravitational potential, perhaps as a result of a
disturbance of the potential by past mergers. While M01 envisioned radial
``sloshing'', \citet{keshet10} offered a more plausible scenario with a
tangential flow of the cool gas beneath the fronts being responsible for
centripetal acceleration.

\citet{ascasibar06} (A06) have reproduced this phenomenon in
high-resolution hydrodynamic simulations of idealized binary
mergers. To explain the absence of the gas disturbance on large
scales, the small infalling subcluster should have no gas (only the
collisionless dark matter --- perhaps having lost its gas during
previous stages of infall), thus disturbing only the gravitational
potential of the main cluster without generating shocks in the
gas. Such a disturbance results in a displacement between the gas peak
and the collisionless dark matter peak, which sets off sloshing, which
continues for several billion years, generating concentric cold fronts
in a spiral pattern (if the merger had any angular momentum), as is
often observed in cool cores. All that is required is a radial
gradient of the specific entropy (which is always present in relaxed
clusters) and an initial gas displacement. The idea that this class of
cold fronts is the result of oscillations of the dark matter peak
caused by a merger has been first proposed by \citet{tittley05}; the
initial displacement may also be caused by the passage of a mild shock
\citep[][]{churazov03,fujita04}.  Sloshing fronts are easily detected
in cool cores, often delineating their boundary, but they are not
confined to cool cores --- in A06, a sufficiently strong initial
disturbance caused detectable fronts to propagate to large radii.
Indeed, cold fronts of this nature (that is, not associated with any
infalling subclusters) have recently been found at $r\sim 0.7-1$ Mpc
in Perseus \citep{simionescu12} and in A\,2142 \citep{rossetti13}, see
Fig.\ \ref{fig:oph_a2142}b.Note that this does not mean that the
cluster gas oscillates from the center all the way out to those large
distances.  While sloshing does begin as a physical displacement of
the gas density peak from the potential peak, the fronts propagate
outwards as waves --- see Fig.\ 8 in A06 and \citet{nulsen13}, who
describe sloshing and the resulting cold fronts as superposition of
g-mode oscillations in the stratified cluster atmosphere.

Core sloshing has a number of important effects on clusters. One is
seen in the X-ray image of the Ophiuchus cluster (Fig.\
\ref{fig:oph_a2142}a; a matching snapshot from the numeric simulations
can be seen in Fig.\ 7c of A06). The gas density peak (which contains
the lowest-entropy gas of the cool core) is completely displaced from
the center of the cD galaxy (shown by a cross). This would temporarily
starve the central AGN of its fuel and may stop its activity.
\citet{zuhone10} showed that sloshing can also facilitate heat
transport from the hot reservoir outside the core into the cool core
via mixing, which can compensate for most of radiative cooling in the
core, provided that gas mixing is not suppressed.  \citet{keshet10}
and \citet{zuhone11} showed that tangential velocity shear in a
sloshing core should amplify and reorder the initially tangled
magnetic field. This effect has observable consequences for the shapes
of cold fronts, which offers an independent tool to study those
magnetic fields, as discussed below. There is also a physical
connection between gas sloshing and cold fronts on one side and the
radio-emitting ultrarelativistic electrons in the cluster cores on the
other (\citet{giacintucci14}; \citet{zuhone13b}).

\begin{figure*}[t]
\centering
\includegraphics[width=0.7\linewidth]%
{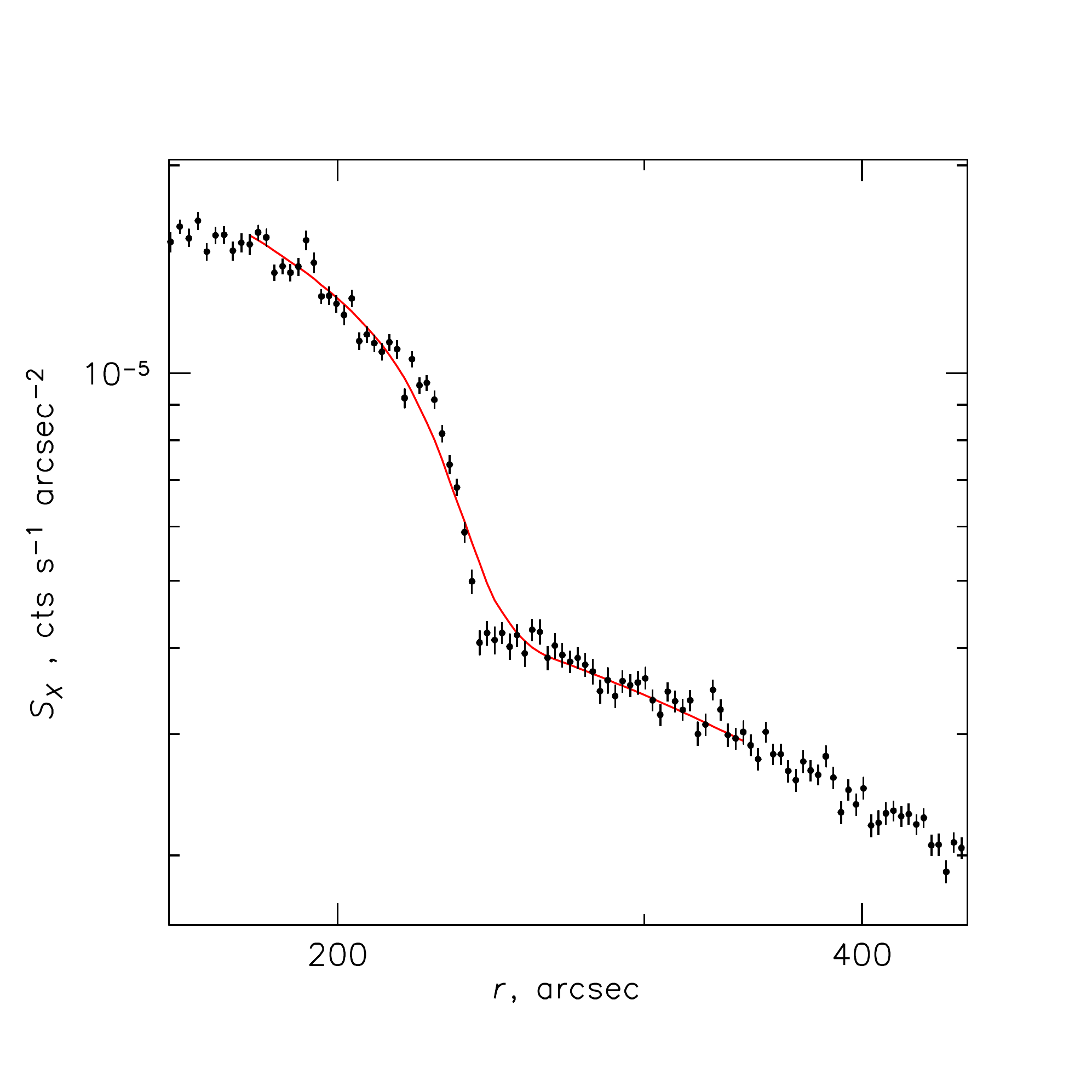}
\caption{X-ray surface brightness profile across the cold front in A\,3667
  (V01, MV07). Red line shows a best-fit model of a density jump that is
  smoothed with $\sigma=11$ kpc, which is the m.f.p.\ for Coulomb diffusion
  from the dense side to the less dense side of the front. If diffusion were
  present, the front would have been smeared by several times this width;
  such diffusion is clearly excluded by the data.}
\label{fig:a3667_prof}
\end{figure*}

\begin{figure*}[t]
\centering
\includegraphics[width=1.0\linewidth]%
{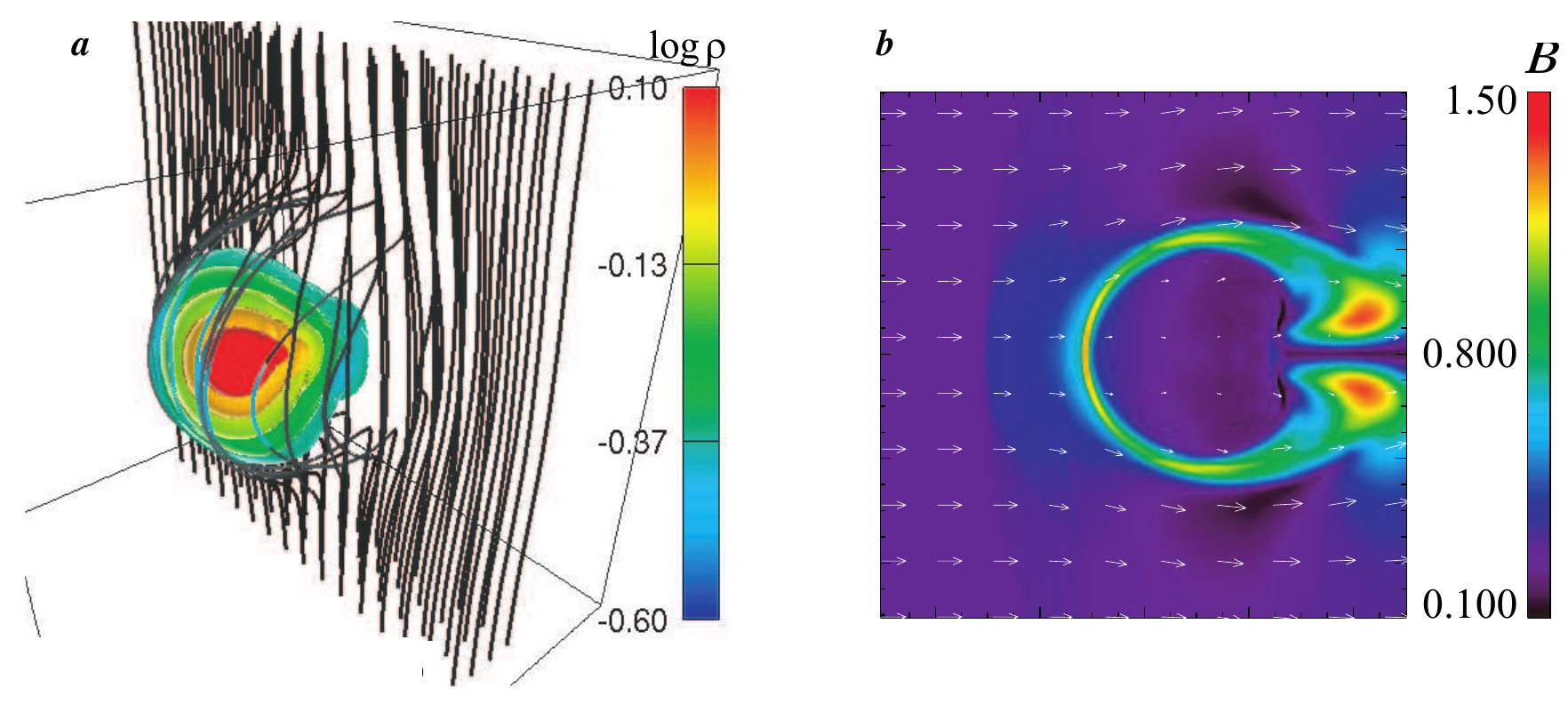}
\caption{Magnetic field draping around an infalling
  subcluster (panel {\em a}). This results in a layer of amplified and
  ordered field immediately outside the cold front (panel {\em b}). The
  field is initially uniform, but the effect for a tangled field is similar.
  \citep[Reproduced from][]{asai05}}
\label{fig:asai_draping}
\end{figure*}

\begin{figure*}[t]
\centering
\includegraphics[width=1.0\linewidth]%
{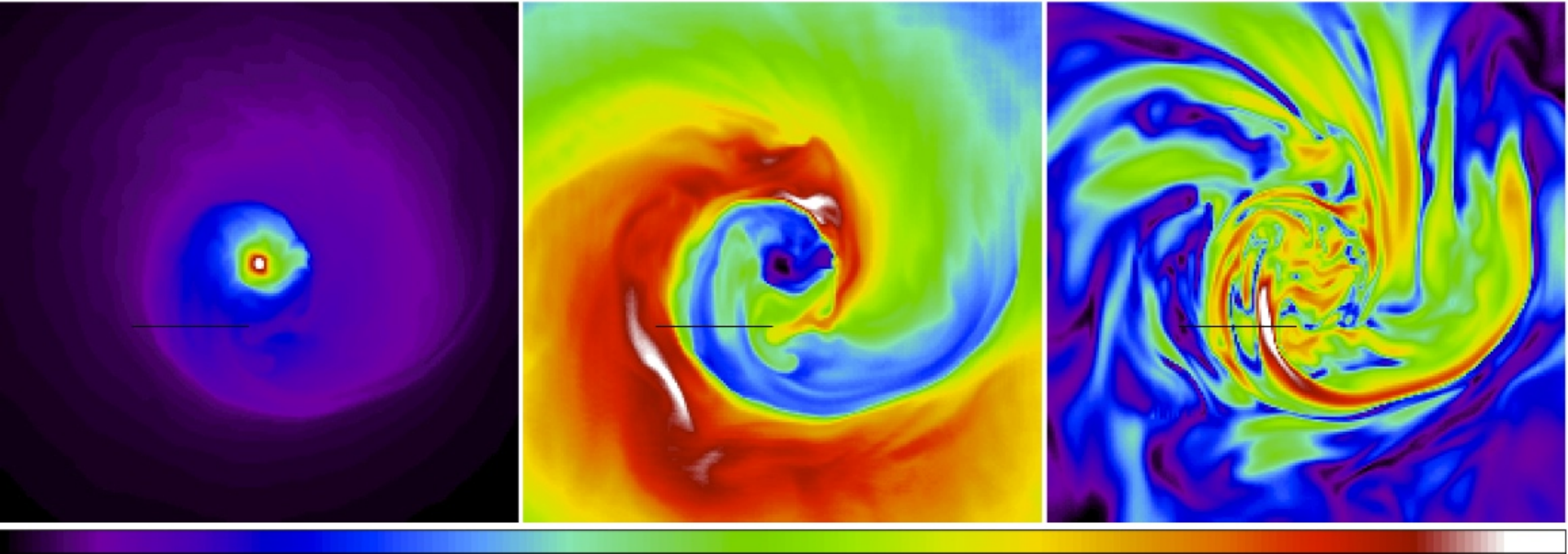}
\caption{The magnetic field around a ``sloshing'' cold front. Left panel
  shows X-ray surface brightness, middle panel shows gas temperature, and
  right panel shows the magnetic field strength. Unlike for the
  ``stripping'' front (Fig.\ \ref{fig:asai_draping}, the field is strongly
  amplified {\em inside}\/ the front \citep[Reproduced from][]{zuhone11}.}
\label{fig:zuhone_slosh_b}
\end{figure*}

\begin{figure*}[t]
\centering
\includegraphics[width=0.4\linewidth,bb=4 280 276 552,clip]%
{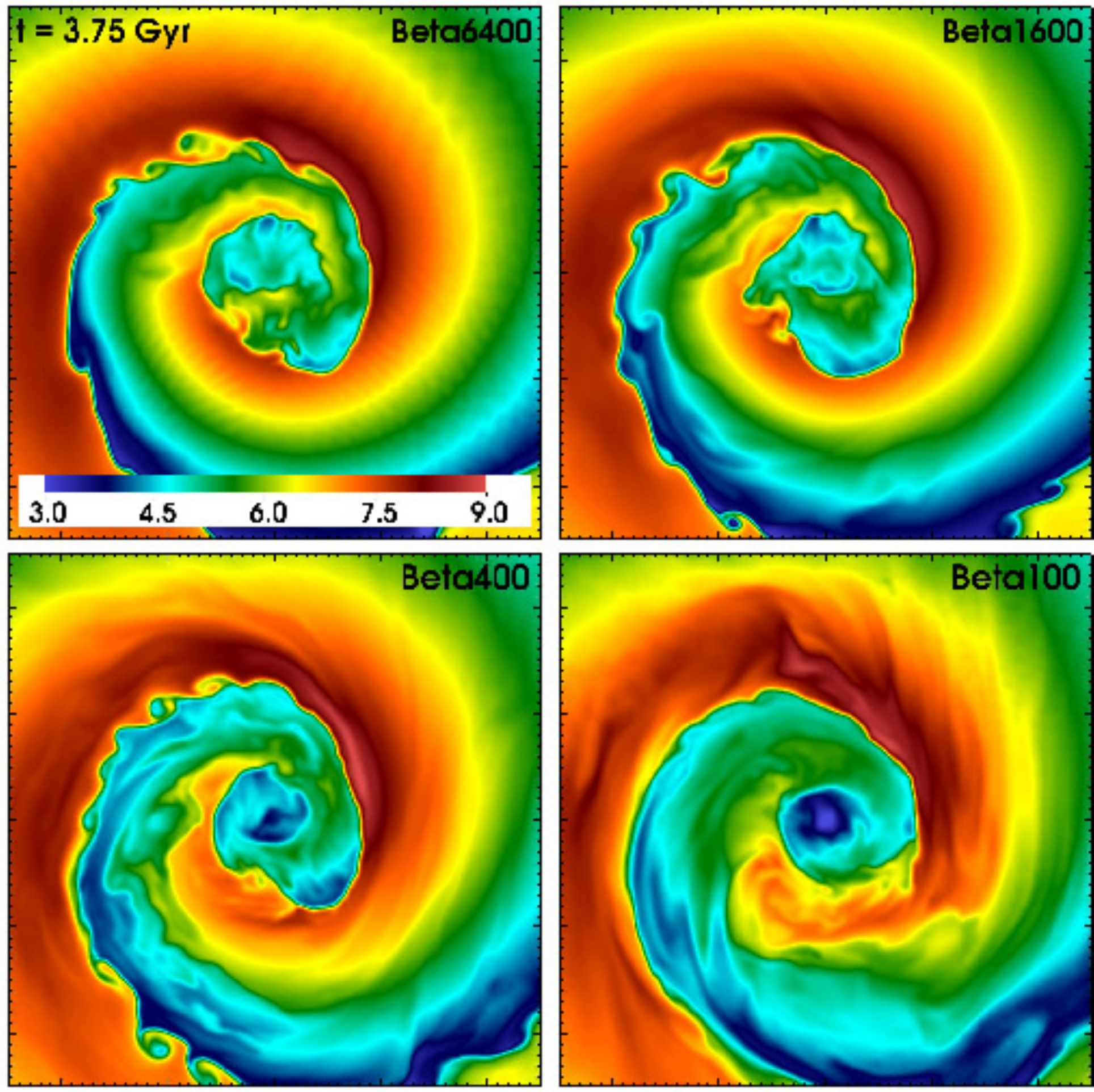}
\includegraphics[width=0.4\linewidth,bb=284 2 556 274,clip]%
{zuhone_cf_b_fig20.pdf}
\caption{The stabilizing effect of the magnetic field on cold
  fronts. Panel size is 500 kpc; color shows gas temperature (the scale is
  in keV). Left panel shows a simulation with a weak field (initial plasma
  $\beta=6400$), right panel shows a stronger, more realistic field
  ($\beta=100$). The realistic field suppresses K-H instabilities, leaving
  relatively undisturbed cold fronts, similar to those observed. (Reproduced
  from ZuHone et al.\ 2011.)}
\label{fig:zuhone_b_kh}
\end{figure*}

\begin{figure*}[t]
\centering
\includegraphics[width=0.9\linewidth]{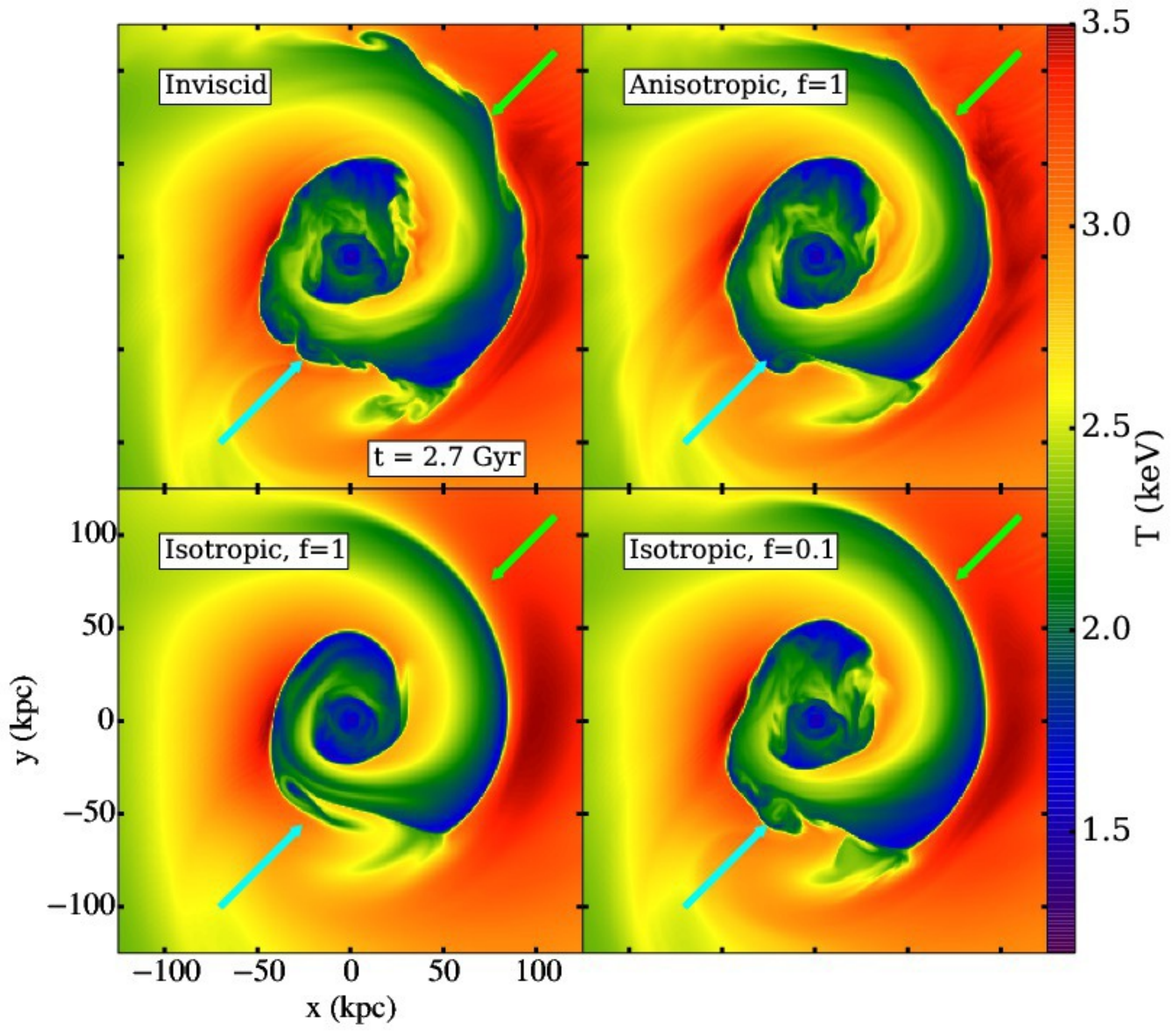}
\caption{Simulation of a sloshing core with different physical models for
  plasma viscosity: inviscid (with a weak magnetic field present), isotropic
  Spitzer viscosity, anisotropic Braginskii viscosity along the field lines,
  and isotropic viscosity suppressed by factor $f=0.1$. A highly suppressed
  viscosity is required to produce visible K-H eddies at cold fronts.
  (Reproduced from ZuHone et al.\ 2014.)}
\label{fig:zuhone_visc}
\end{figure*}

\subsection{Physics of cold fronts}

Cold fronts can be used to study the microphysics of the intracluster
plasma.  One property of the cold fronts observed with Chandra that is
immediately striking is their sharpness. Fig.~\ref{fig:a3667_prof}
shows an X-ray brightness profile across the prominent front in
A\,3667 (Fig.~\ref{fig:a3667_1e}a).  \citet{V01} pointed out that the
front is sharper than the mean free path for Coulomb
collisions. Indeed, red line in Fig.~\ref{fig:a3667_prof} shows a
best-fit model that includes broadening of the front with a width
equal to the Coulomb m.f.p. for diffusion from the cool inner side to
the hot outer side of the front. Such broadening is clearly
inconsistent with the data, which means the diffusion across the front
is suppressed. \citet{ettori00} pointed out that the existence of the
observed temperature jumps at cold fronts implies that thermal
conduction is also strongly suppressed. \citet{vikhlinin01a} proposed
(within the ``subcluster stripping'' scheme of M00) that the motion of
the infalling subcluster through a tangled magnetic field frozen into
the ambient gas of the main cluster would naturally form an insulating
layer of the field oriented strictly along the front surface, as a
result of the field ``draping'' around an obstacle, as originally
proposed by Alfv\'en to explain the comet tails. This effect is shown
in a simulation by \citet{asai05} in Fig.\ \ref{fig:asai_draping}. As
pointed out by \citet{lyutikov06}, such draping can amplify the field
in the narrow layer immediately outside the front to values
approaching equipartition with thermal pressure (compared to magnetic
pressures of order 1\% of thermal pressure in the rest of the
cluster). Such a layer would completely suppress diffusion and thermal
conduction across the front, while the magnetic tension of such a
layer may stabilize the front against Kelvin-Helmholtz instability
(\citet{vikhlinin01a}, MV07), which we will see below in a simulation.

If a cluster with a cool core initially has a tangled magnetic field,
when the core is disturbed and starts sloshing, the field is rapidly
stretched by the tangential gas velocity shear \citep{zuhone11},
forming layers of amplified field oriented along the surface of the
cold front form.  These layers are located {\em under}\/ the front
surfaces, unlike for the draping effect in a ``stripping'' front. A
snapshot from the simulations of this process is shown in Fig.\
\ref{fig:zuhone_slosh_b}. Such layers can suppress KH instability of
the front surface (Fig.\ \ref{fig:zuhone_b_kh}) similarly to a draping
layer. However, there is an important difference for the effective
thermal conduction across the front.

If a subcluster falls into a cluster from a large distance, it will be thermally
insulated from the ambient gas by a magnetic draping layer for as long as the
subcluster survives as a coherent structure, because the initially disjoint
magnetic field structures of the cluster and subcluster cannot connect at any
stage of the infall (as long as there is no magnetic reconnection) --- even when
the gases are geometrically separated only by a thin cold front. Since heat is
conducted only along the field lines, it is not surprising to see a cool
infalling subcluster survive the immersion into the hot gas of the bigger
cluster \citep[e.g., the group falling into A2142 discovered by][]{eckert14}.
However, for a sloshing cold front, thermal conductivity across the front may
not be completely suppressed by the magnetic layer, as shown by simulations of
\citet{zuhone13a}). The reason is that prior to the onset of sloshing, the
regions inside and outside the cluster core are connected by the field lines,
and while sloshing stretches most of them tangentially, it does not sever the
connection completely. Those authors suggested that the existence of the
temperature jumps across the sloshing cold fronts may therefore be used to
constrain heat conduction {\em along}\/ the field lines.

Another interesting (and completely unknown) plasma property that may
be constrained by the observations of cold fronts is viscosity (MV07).
Fig.~\ref{fig:zuhone_visc} \citep[from][]{zuhone14} shows a simulated
sloshing core with the plasma viscosity modeled in different ways. The
viscosity, either isotropic or anisotropic (it is likely to be the
latter in the presence of magnetic fields), acts to suppress the KH
instabilities, and the effect should be observable. Of course, the
effect of viscosity is superimposed on the stabilizing effect of the
magnetic layers discussed above, so the observations will most likely
be able to constrain some combination of the two. Based on simulations
without the magnetic field, \citet{roediger13} concluded that the
effective isotropic viscosity should be significantly suppressed to
explain the disturbed appearance of a cold front in Virgo.

This area of research is currently under rapid development, with
high-quality observations of cold fronts being obtained and tailored numeric
simulations being constructed, so interesting constraints on the
microphysical properties of the cluster plasma can be expected soon.



\section{Shock fronts and non-thermal components}

\subsection{Shock fronts}

Shock waves in clusters of galaxies are the main agents to convert the
kinetic energy of supersonic and superalfv\'enic plasma flows produced
by gas accretion, merging substructures and AGN outflows into both the
thermal and the non-thermal components. The shocks are essential to
heat the gas and to produce the observed thermal and non-thermal
radiation.  The shock structures in plasma and their ability to create
the non-thermal components -- energetic charged particles and
electromagnetic fields -- depends on whether the shock is collisional
or collisionless. The mean free path of a proton due to the Coulomb
collisions is $\lambda_{\rm p} \approx 7\times 10^{21} v_8^4
n_{-3}^{-1}$ (here $v_8$ is proton velocity in thousands \kms and
$n_{-3}$ is the ambient gas number density in 10$^{-3}$~\cmc). Note
that an ion of charge $Z$ and atomic weight $A$ have a gyroradius
$r_{\rm gi} = 3.3 \times 10^9 (A T_{\rm keV})^{1/2} (Z B_{\mu {\rm
    G}})^{-1}$ cm in a magnetic field $B_{\mu {\rm G}}$ measured in
$\mu$G. The Coulomb mean free path is much larger than the proton
gyroradius (as well as the ion inertial length which we shall
introduce later) at all particle energies of interest in the cluster
and thus the plasma shocks in clusters are expected to be
collisionless as it is the case in the hot interstellar and
interplanetary plasmas.

A specific feature of the collisionless shocks is the mechanism of the
flow momentum and the energy dissipation by means of the excitation
and damping of collective electromagnetic fluctuations providing
numerous degrees of freedom with a very broad range of the relaxation
times and spatial scales. Moreover, the collisionless shock may
accelerate particles to ultra-relativistic energies resulting in a
broad particle energy spectra formation. The multi-scale nature of the
shock with a strong coupling between the scales makes the problem of
laboratory studies and theoretical modeling of the structure of the
collisionless shock to be very complicated.  However, some basic
features of the collisionless shock physics were established from
direct interplanetary plasma observations, imaging and spectroscopy of
shocks in supernova remnants and computer simulations \citep[see
e.g. the review by][]{treumann09}.

The shock flow dissipation mechanisms depend on the flow velocity,
magnetization and shock obliquity.  Shocks of low enough Mach numbers
${\cal M}_{\rm s} < {\cal{M}_{\rm crit}}$ are able to dissipate the
kinetic energy of the flow by anomalous Joule dissipation. In the lack
of Coulomb collisions the dissipation is usually associated with the
anomalous resistivity provided by wave-particle interactions. Such
shocks are called subcritical and $\cal{M}_{\rm crit}$ is dubbed the
first critical Mach number \citep[see e.g.][]{kennel85}. The structure
of the shock transition in this case may be smooth and laminar, but
this mainly occurs for low $\beta$ shocks.  The magnetization
parameter $\beta = 8\pi nT/B^2 = {\cal M}^2_{\rm a}/{\cal M}^2_{\rm s}
\approx 40\, n_{-3}\, T_{\rm keV}\, B_{\mu G}^{-2}$, characterizes the
ratio of the thermal and magnetic pressures. The first critical Mach
number is maximal for a quasi-transverse shock in the plasma with low
magnetization $\beta >$ 1 and it is below 2.76. At large $\beta$
typical for the hot intracluster plasmas some transverse shocks as
well as very weak shocks can be subcritical, while most of the shocks
are supercritical. The width of the viscous transition in the
collisionless subcritical shock wave can be estimated as $l_{\rm
  e}/\sqrt{{\cal M}_{\rm s} - 1}$, where $l_{\rm e} = c/\omega_{\rm
  pe} \approx 1.7 \times 10^7 n_{-3}^{-0.5}$ cm. Here $\omega_{\rm
  pe}$ is the electron plasma frequency.

\begin{figure*}[t]
\centering
\includegraphics[width=0.9\linewidth]{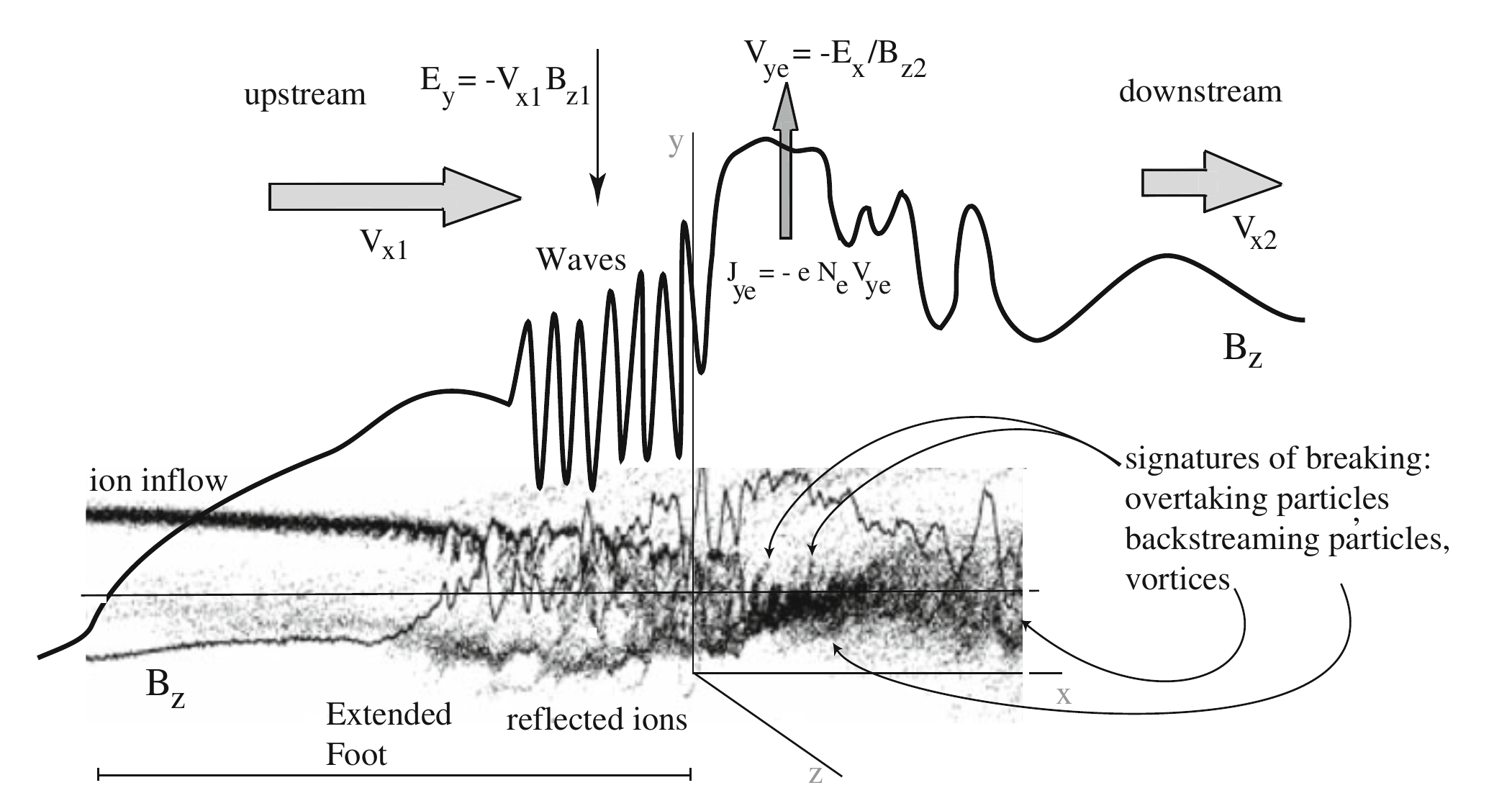}
\caption{Schematic structure of a perpendicular supercritical shock. The
 profiles of the magnetic field and plasma density at the shock transition
 region dominated by the ion reflection dissipation mechanism are shown.
 Charge separation over an ion gyro-radius  in the shock ramp magnetic field
 produce the electric field along the shock normal which may reflect the
 incoming ions back upstream. The electric field along the shock front produced
 by conductive plasma motion in the magnetic field may accelerate particles by
 the electric drift acceleration mechanism. The magnetic field of the current
 carried by the accelerated back-streaming ions causes the magnetic foot in
 front of the shock ramp. Reproduced from \citet{treumann09}.}
\label{fig:sketch}
\end{figure*}

If the shock Mach number exceeds $\cal{M}_{\rm crit}$ the anomalous
resistivity is unable to provide the required dissipation rate to
satisfy the Rankine-Hugoniot conservation laws at the shock. Then the
shock became supercritical and its structure is turbulent. The
dissipation mechanism of a supercritical shock is usually dominated by
ion reflection. Some fraction of the incoming ions are reflected by a
force which is a combination of electrostatic and magnetic fields and
the reflected ion behavior depends on angle between the local upstream
magnetic field and the local shock normal. In Fig.~\ref{fig:sketch},
taken from \citet{treumann09}, we illustrate the basic processes in
the collisionless perpendicular shocks.

The width of the supercritical shock transition in case of a
quasi-parallel shock may reach a few hundreds ion inertial lengths
which is defined as $l_{\rm i} = c/\omega_{\rm pi} \approx 7.2 \times
10^8 n_{-3}^{-0.5}$ cm. Here $\omega_{\rm pi}$ is the ion plasma
frequency and $n$ is the ionized ambient gas number density measured
in \cmc. The shock transition structure is unsteady in the case of
magnetized shocks being a subject of shock front reformations which however
are expected to be suppressed in the hot intercluster plasmas.  The widths
of viscous transition in the collisionless shocks is orders of
magnitude less the Coulomb mean free path and it is below the spatial
resolution of optical and X-ray observations even for a few kpc
distance galactic supernova shocks.

Non-thermal energetic electrons are responsible for most of the
observed radio emission from clusters, but the exact nature of
particle acceleration process and the origin of the observed $\sim
\mu$G strength magnetic fields in clusters are still under debate.
From the energetic ground shock waves are the most natural
accelerating agent in hot weakly magnetized plasmas. As we have
learned from multi-wavelength observations of strong shocks in young
supernova remnants in order for diffusive shock acceleration (DSA) to
be fast enough to reach 100 TeV regime particle energies observed in the
sources, significant non-adiabatic magnetic field amplification is
required \citep[see e.g.][]{bell04,amato09,bykov12,schure12}. The
cosmic ray driven instabilities may provide a source of free energy
for strong magnetic field amplification transferring a few percent of
ram pressure of a strong shock (with ${\cal M}_{\rm s} \gg 1$) into
fluctuating magnetic fields \citep[see Fig. 11 in][]{beov14}. This is
indeed enough to provide $\mu$G -- magnetic fields behind the strong
large scale accretion shock. The observed $\mu$G magnetic field
strength is well above the field amplitude produced by the adiabatic
compression of the intergalactic field at an accretion
shock. Therefore it is possibly produced by CR driven
instabilities at the strong accretion shock.  The field amplification
mechanism driven by instabilities of the anisotropic CR distribution
produced by DSA is efficient if CRs get a substantial fraction of the
shock ram pressure, which is expected in strong shocks with hard CR
spectra.  In the case of strong amplification of the fluctuating
magnetic field in the shock precursor, the direction of the local
magnetic field just before the viscous plasma shock transition will
vary with time providing periods of both quasi-perpendicular and
quasi-parallel configurations. This may differ from the case of weak
shocks with steep spectra of accelerated particles which may not
contain enough free energy in the high energy end of particle spectra
to provide strong long-wavelength turbulence in the shock upstream.

Because of much higher pre-shock densities, the internal shocks with
modest Mach numbers ${\cal M}_{\rm s} <$4 in clusters, which are
propagating in a very hot intracluster plasma, dissipate more energy
than the strong accretion shocks.  Internal shocks with 2$< {\cal
  M}_{\rm s} <$4 were shown in simulations by \citet{ryu03} to produce
about a half of the total kinetic energy dissipation, while the
internal shocks as a whole are responsible in this model for about
95\% of gas thermalization. Microscopic simulations of electron
acceleration in quasi-perpendicular shocks of ${\cal M}_{\rm s} <$5
were performed by \citet[][]{guoea_el_accel14}. This particle-in-cell
plasma modeling demonstrated that the repeated cycles of shock drift
acceleration (SDA) may form power-law electron energy spectra.  In a
particular run with a quasi-perpendicular pre-shock magnetic field and
${\cal M}_{\rm s}$ = 3 they found that about 15\% of the electrons
were accelerated and a power-law electron energy spectrum with a slope
of $q \approx$ 2.4 was formed. The energy density carried by the
accelerated energetic electrons in this simulation was about 10\% of
the bulk kinetic energy density of the incoming ions. This electron
acceleration efficiency is much higher than that estimated from
supernova remnant observations. The transverse box size in this
simulation was  about one ion gyro-radius. It is important to
confirm the interesting result with larger simulation boxes  since the
shock structure is determined by ions and the model have to fulfill
the requirements of the theory of charged particle motion in an
electromagnetic field with one ignorable spatial coordinate by
\citet{jones98}.

\begin{figure*}[t]
\centering
\includegraphics[width=0.9\linewidth]{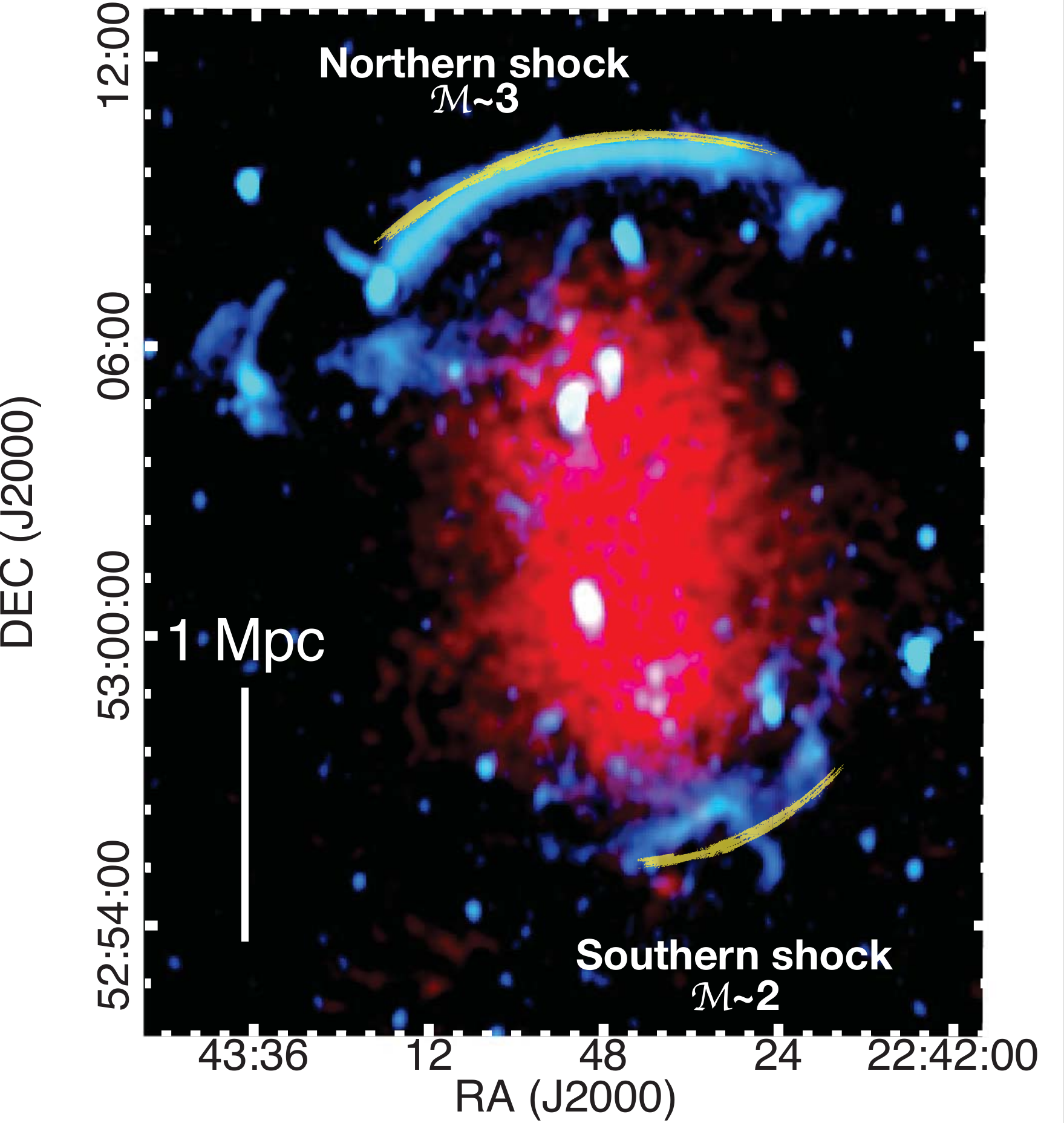}
\caption{Smoothed 0.5-2.0 keV band X-ray image (red) and WSRT 1.4 GHz image of
CIZA J2242.8+5301 (cyan). The thin yellow lines depict
the approximate locations of the shock fronts confirmed by Suzaku.
The figure is courtesy  H.Akamatsu et al. (A\&A, v.582, id.A87, 2014).
}
\label{fig:CIZA}
\end{figure*}

The very long (Mpc scale) highly polarized radio structure observed in
the merging galaxy cluster CIZA J2242.8+5301 by \citet{vweeren10} can
be understood in this way, as the associated weak shock may not
disturb the initial inter-cluster magnetic field in the shock
upstream. The observed polarization favors a transverse shock
configuration which may provide efficient shock drift acceleration of
relativistic electrons \citep[][]{guoea_el_accel14,guo14a}.  This would
require a very uniform magnetic field in the shock upstream.  The
radio structures and the X-ray emission of CIZA J2242.8+5301 cluster
illustrating the location and the extension of the large scale shocks
of moderate strengths are shown in Fig.~ \ref{fig:CIZA} from
H.Akamatsu et al. (A\&A, v.582, id.A87, 2014).

\subsection{Intracluster cosmic rays}

Both electrons and ions are likely accelerated at the shock fronts
produced by mergers or supersonic outflows, but their subsequent
evolution differ markedly.  Since the ultra relativistic electrons
have a radiative lifetime much shorter than the age of the cluster,
they rapidly radiate most of their energy away and then may comprise a
long lived population at Lorentz factors of around 100 where both
Coulomb and radiation losses are longer than 10$^9$ years
\citep{petrosian08}.  On the other hand, ions (protons mostly) lose
only a small fraction of their energy during the lifetime of the
cluster, and their diffusion time out of the cluster is even larger,
so that they are stored in clusters for a cosmological time scale
\citep{voelk96,berezinsky97}.

The long lived non-thermal particle populations may be the subject of
a consequent re-acceleration by multiple shocks of different strengths
as it is likely the case in galactic superbubbles and starburst
regions \citep[c.f.][]{bykov01,AARv14}.  Relativistic electrons can be
re-accelerated in-situ by a long-wavelength MHD turbulence
\citep{bt93} which may be generated in the ICM during cluster mergers
\citep{2011MNRAS.410..127B,2014IJMPD..2330007B}.  The energy contained
in both populations depends on the energy spectra of particles.
Electrons of energies at about 50 MeV are difficult to constrain from
the observational point of view. Gamma-ray observations are used to
constrain the energy density in relativistic particles \citep[see
e.g.][]{Fermi_clusters14}.  Recently, \citet[][]{prokhorov_churazov14}
analyzed {\sl Fermi-LAT} photons above 10 GeV collected from the
stacked 55 clusters selected from a sample of the X-ray brightest
clusters.  They obtained an upper limit of the pressure of
relativistic protons to be ~1.5\% relative to the gas thermal energy
density, provided that the spectral index $q$ of relativistic proton
power-law distribution is 2.1, while for $q$=2.4 the limit is already
about 6\%. These estimations assume that relativistic and thermal
components are mixed. Similar results were reported by
\citet[][]{Fermi_clusters14} who analyzed another set of clusters at
the photon energies starting from 500 MeV.  The observations reported
above have placed stringent limits on the pressure contained in the
high-energy particle populations with hard spectra, while the
constraints on the non-thermal components with steep enough spectra to
be dominated by the sub-GeV particles remain to be established.

\begin{figure*}[t]
\centering
\includegraphics[width=0.9\linewidth]{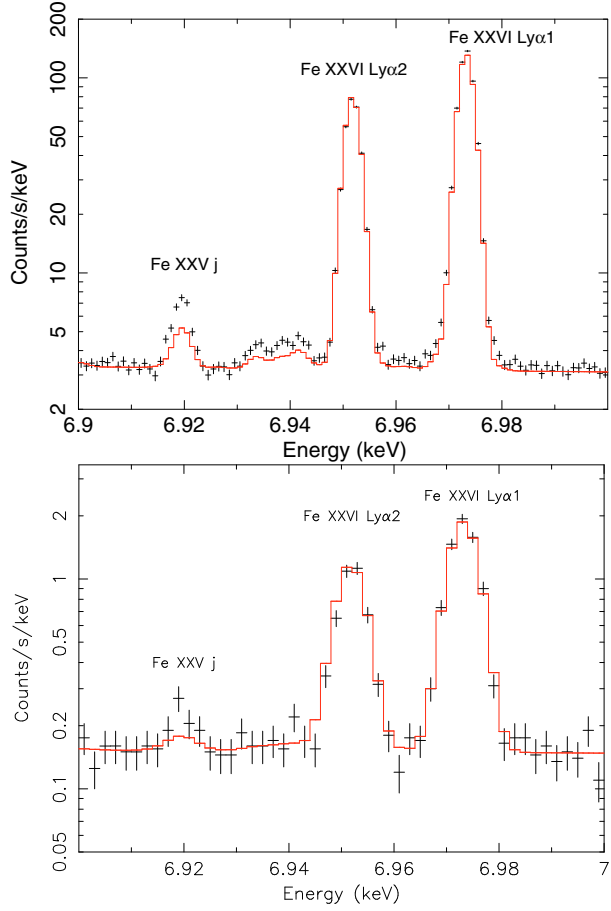}
\caption{X-ray micro-calorimeter spectra from a 1\arcmin\ region at the center of
a bright cluster with a mixture of the thermal and supra-thermal electrons as
simulated by \citet{kbw09}. Simulations made for 100 ks observations with
Athena (top panel) and ASTRO-H (bottom panel). The energy distribution of
electrons  accelerated at ${\cal M}=2.2$ Mach number shock was calculated with the
kinetic electron acceleration model by \citet{bu99}. Crosses are simulated data
points. Solid line: best-fit model to a pure Maxwellian plasma, with temperature
16.99 keV. Note the excess emission of satellite lines in the data, in
particular the Fe XXVI-line.}
\label{fig:Fe_sat_lines}
\end{figure*}

The most direct evidence of a non-thermal component within the
intracluster medium comes fram radio observations, that have revealed
the presence of two main classes of Mpc-scale diffuse synchrotron
sources, generally called ``radio halos'' and ``radio relics''
\citep[see e.g.][]{ferrari08,feretti12}. The origin of cosmic rays in
radio halos is still elusive, even though their existence has been
known for decades. The morphological, spectral and polarization
properties of the radio halos were attributed by \citet{Brunetti01} to
radiation of relativistic electrons accelerated by ICM
turbulence. Alternatively, the radiating leptons could be of the
secondary origin i.e. a product of hadronic collisions of
(long--lived) non-thermal energetic protons with thermal nuclei
\citep{1980ApJ...239L..93D,1999APh....12..169B}.  In this case, one
should also expect $\gamma$-ray emission, which has not been seen so
far. MHD simulations carried out by \citet{2010MNRAS.401...47D}
indicated some problems in explaining the observed steepening of the
synchrotron spectra of halos. However, hadronic models with
energy-dependent CR proton transport coefficients may be able to
reproduce the steep spectra \citep{2011A&A...527A..99E}.

Radio relics are considered to be related to DSA, even if some of
current observational results suggest the need to review or refine
electron acceleration models within this class of sources
\citep{2012A&A...543A..43V, 2014MNRAS.443.2463O}. Radio observations
of most kinds of relic sources in clusters of galaxies show rather
steep synchrotron spectra ($S_{\nu} \propto \nu^{-\alpha}$), with
$\alpha \gtrsim 1$ \citep[see e.g.][]{miley80,ferrari08,feretti12}.
As expected in DSA models, the spectral index $\alpha$ can change
within the source, with a steepening from the front towards the back
of the shock (see e.g. the spectacular radio relic in the cluster CIZA
J2242.8, whose spectral index is ranging from 0.7 to about 1.7;
\citet[][]{stroeAA13}). The corresponding indices of the relativistic
lepton power-law spectra are $q \gtrsim 2.5$. The soft particle spectra are
expected to be produced by relatively weak shocks in both the
diffusive shock acceleration model \citep[see e.g.][]{bruegen12} and
in the shock drift electron acceleration model by
\citet[][]{guoea_el_accel14}.  The simulated statistical distribution
of the merging shock strengths \citep[see e.g.][]{ryu03} peaks at
rather weak shocks, and therefore the constrains on the pressure in
relativistic leptons of Lorentz factors of about 100 and the
non-thermal sub- and semi-relativistic protons of soft energy spectra
still need further substantiation.

Apart from radio and gamma-ray observations discussed above fine X-ray
spectroscopy has a potential to study non-thermal components. To
illustrate the effects of such a supra-thermal electron distribution
on data, \citet{kbw09} simulated two X-ray micro-calorimeter spectra
extracted from a circular region with a radius of 1\arcmin\ centered
on the core of a bright cluster with a 0.3--10 keV luminosity of
$6.3\times 10^{37}$~W within the extraction region, at an assumed
redshift of $z = 0.055$. In the simulation of the spectrum of a deep
100 ks X-ray micro-calorimeter observation, they assumed a post-shock
downstream electron distribution for a Mach number of ${\cal M} = 2.2$
and pre-shock temperature of $kT = 8.62$~keV (i.e. 10$^8$ K)
calculated with a kinetic model of electron heating/acceleration by a
collisionless shock developed by \citet{bu99}.  In
Fig.~\ref{fig:Fe_sat_lines} taken from \citet{kbw09} we present the
6.9--7.0 keV part of the spectrum (rest-frame energies) which shows
the Fe XXVI Ly$\alpha$ lines and the Fe XXV j-satellite line simulated
for a 100 ks observation with the X-ray micro-calorimeter aboard the
{\sl ASTRO-H} mission and also for that proposed for the future {\sl
  Athena} mission.  Enhanced equivalent widths of satellite lines are
shown to be a good indicators of non-thermal electrons. The satellite
line in the simulated spectrum is clearly stronger than that predicted
by the thermal model with a Maxwellian electron distribution.  The
energy density in the supra-thermal electrons modeled by
\citet[][]{guoea_el_accel14} is larger than that was used in the
simulations by \citet{kbw09} and therefore it can be observationally
tested in this way.  This illustrates a good potential of
high-resolution spectra obtained by future satellites with a large
effective area to observationally reveal non-Maxwellian tails in the
electron distributions.

\subsection{Intracluster magnetic fields}

The study of the linearly polarized emission from radio images is of
cardinal importance in order to constrain the properties of
intra-cluster magnetic fields.

\citet{2010A&A...514A..71V} studied the power spectrum
  of the magnetic field associated with the giant radio halo in the
  galaxy cluster A\,665. They performed sensitive observations of this
  cluster with the Very Large Array at 1.4~GHz and compared these with
  simulations of random three-dimensional turbulent magnetic fields,
  trying at the same time to reproduce the observed radio continuum
  emission from the halo. They constrained the strength and structure
  of the intracluster magnetic field by assuming that it follows a
  Kolmogorov power-law spectrum and that it is in local equipartition
  with relativistic particles. A central magnetic field strength of
  about 1.3$\mu$G was inferred for A\,665. The azimuthally averaged
  brightness profile of its radio emission suggests the energy density
  of the magnetic field to follow the thermal gas density, leading to
  an averaged magnetic field strength over the central 1~Mpc$^3$ of
  about 0.75$\mu$G. An outer scale of the magnetic field power
  spectrum of $\sim$ 450 kpc was estimated from the observed
  brightness fluctuations of the radio halo.

\citet{2011A&A...530A..24B} used the Northern VLA Sky Survey to
analyze the fractional polarization of radio sources out to 10 core
radii from the centers of 39 massive galaxy clusters with the aim to
find out how different magnetic field strengths affect the observed
polarized emission along different sight lines through the clusters.
They found the fractional polarization to increase towards the cluster
peripheries, and their findings are in accord with a magnetic-field
strength of a few $\mu$G in the centers. A statistical test gives no
hint at any differences in depolarization for clusters with and
without radio halos, but indicates significant differences of the
depolarization of sources seen through clusters with and without cool
cores.

\citet{2013A&A...554A.102G} investigated the potential of
new-generation radio telescopes (e.g. the Square Kilometer Array, SKA
and its pathfinders) in detecting the polarized emission of radio
halos in galaxy clusters. To this end, they used
magneto-hydrodynamical simulations conducted by
\citet{2011ApJ...739...77X,2013MNRAS.435.3575B} to predict the
expected polarized emission of radio halos at 1.4~GHz. The synthetic
maps of radio polarization were compared with the detection limits on
polarized emission from radio halos set by current and upcoming radio
interferometers. They show that both the angular resolution and
sensitivity expected in future sky surveys at 1.4 GHz using the SKA
precursors and pathfinders (like APERTIF, ASKAP, Meerkat) are rather
promising for the detection of the polarized emission of the most
powerful ($L_{1.4 \rm GHz} > 10^{25}$~W~Hz) radio halos. Furthermore,
the upgraded JVLA has the potential to detect polarized emission from
strong radio halos already now, though with relatively low angular
resolution. However, the detection of polarization signal in faint
radio halos ($L_{1.4 \rm GHz} < 10^{25}$~W~Hz) has to await the fully
deployed SKA.


\section{Turbulence in clusters}
\label{sec:3.2.3}

Turbulence and density fluctuations affect the basic assumptions
behind hydrostatic equilibrium and hence the estimates of important
cosmological parameters such as cluster masses. For this reason, but
even more because of the interesting astrophysical processes involved,
we discuss in more detail the various observational signatures of
turbulence and density inhomogeneities.

We consider here constraints from measurements in the radio band and
pressure maps, surface brightness distributions, resonant scattering
and line broadening in the X-ray range.


\subsection{X-ray view: pressure and surface brightness mapping}

Large scale plasma motions and turbulence in clusters primarily concerns
fluctuations in the velocity field. However, they also lead to density and
pressure fluctuations that can be measured with X-ray  telescopes with
sufficient spatial resolution and only modest spectral resolution (to measure
temperatures).

\citet{2004A&A...426..387S} described this method in a seminal paper on the Coma
cluster. From a mosaic of XMM-Newton observations of Coma, spatial scales
between 20~kpc and 2.8~Mpc could be sampled. Their maps show -- superimposed on
the normal radial density and pressure gradients -- fluctuations up to 2\% in
temperature and 5\% in intensity. These fluctuations are correlated and agree
with the expected adiabatic fluctuations of the pressure. The spatial power
spectrum of these pressure fluctuations is in agreement with a Kolmogorov-like
spectrum.  They found that between 10--25\% of the total intracluster
pressure is in the form of turbulence.

More recently, \citet{2012MNRAS.421.1123C} extended this work by also
including higher spatial resolution observations of Coma with
Chandra. Contrary to \citet{2004A&A...426..387S}, they consider
density fluctuations. They find that the relative density fluctuations
in Coma have amplitudes of 5 and 10\% at scales of about 30 and
300~kpc, respectively. They also consider several other explanations
for the observed density fluctuations. For instance, at large scales,
gravitational perturbations due to the large cD galaxies and entropy
variations due to infalling cold gas may produce apparent density
fluctuations.

\subsubsection{Surface brightness profiles}

In the cluster outskirts, models predict that the X-ray emission
should become more clumpy due to the shallow gravitational potential
and the ongoing strong cluster evolution associated, for instance,
with infall of galaxies or groups towards the growing mass
concentrations. One may define a cluster clumping factor $C$ as

\begin{equation}
C^2 = <n^2> / <n>^2,
\end{equation}

\noindent with $n$ the gas density and the brackets denoting averaging over
volume.

\citet{2011Sci...331.1576S} used Suzaku observations of the outskirts
of the Perseus cluster to determine the cluster properties around the
virial radius and beyond. They find an apparent baryon fraction higher
than the cosmic value at these large radii, which they reconcile by
assuming a very clumpy medium at these distances.

Interestingly, \citet{2013A&A...551A..22E}, by combining pressure
measurements from Planck with density profiles obtained from ROSAT,
found a much lower clumping factor $C$ of about 1.2 at $R_{200}$,
compared to 3--4 by \citet{2011Sci...331.1576S}. These differences may
be associated to the precise way of averaging used in determining $C$.

\subsubsection{Direct measurements of line broadening}

\begin{figure}[t]
\begin{center}
\includegraphics[width=0.7\textwidth, angle=-90]{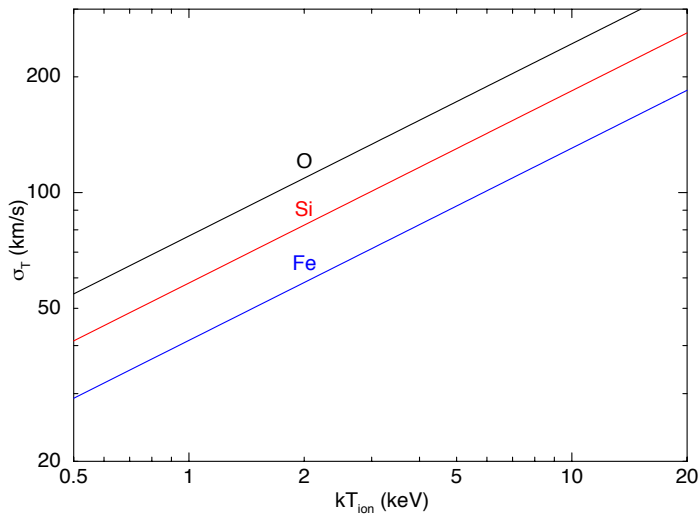}
\end{center}
\caption{Thermal velocity dispersion in the line of sight corresponding to the
thermal motion of emitting ions, for oxygen silicon and iron ions, as a function of gas temperature.}
\label{fig:turbo}
\end{figure}

The most direct way to measure turbulence in clusters is to measure the spectral
line broadening caused by the turbulence. Motion of the line-emitting ions will
produce Doppler shifts. However, also thermal motion of the ions produces
Doppler shifts. The width $\sigma_T$ for thermal motion of the ions (for a
Gaussian velocity profile $\sim \exp (-\Delta v^2/2\sigma_T^2)$) is given by

\begin{equation}
\sigma_T/c = 0.00103 T_{\rm{ion}}/\sqrt M_{\rm{ion}},
\end{equation}

\noindent where the ion temperature $T_{\rm{ion}}$ is expressed in keV
and the ion mass $M_{\rm{ion}}$ in atomic units. Thus, for hydrogen
and an ion temperature of 1~keV the thermal broadening is larger than
the turbulent broadening if the turbulent velocity dispersion in the
line of sight $\sigma_{\rm v}$ is less than 300~km\,s$^{-1}$. For the
more relevant case of iron (atomic mass 56.25) the thermal velocity
dispersion is always less than 200~km\,s$^{-1}$
(Fig.~\ref{fig:turbo}), and hence turbulence is relatively easily
detected provided that the detector has sufficient energy resolution
to measure the line broadening.

At this moment the only instrument capable of measuring line
broadening in clusters is the RGS reflection grating spectrometer on
board of XMM-Newton.  However, because this is a slit-less dispersive
instrument, the spatial extent of a cluster gives an additional line
broadening effect, and this must be modeled properly. Only some cool
core clusters have sufficiently small size to allow for such a
study. In addition, there is some uncertainty in the calibration of
the instrumental line spread function. With these caveats in mind,
\citet{2011MNRAS.410.1797S} studied a sample of 62 clusters, groups
and individual ellipticals. They found upper limits for the turbulent
velocity of 200~km\,s$^{-1}$ for five of their studied cases, and one
case (Klemola~44) with a measured broadening of
1500~km\,s$^{-1}$. Optical observations show this latter cluster to be
a very disturbed system. For the sample as a whole, 15 sources appear
to have less than 20\% of their thermal energy in the form of
turbulence.

Clearly, better measurements are needed. ASTRO-H (expected launch
2015) will contain a calorimeter with about 6~eV spectral resolution
(300~km\,s$^{-1}$ at the iron lines), that does not suffer from the
spatial blurring inherent to the RGS. Turbulent broadening of spectral
lines is expected to be measurable down to 50--100~km\,s$^{-1}$,
ultimately limited by the calibration of the instrument.  Measuring
cluster turbulence is one of the key scientific questions that this
satellite will address.

While ASTRO-H has sufficient spectral resolution to measure the
turbulence, its spatial resolution of the order of an arcmin will
limit the amount of mapping that is possible. In the longer future,
the proposed ESA mission Athena (launch 2028) will have a
significantly higher spatial resolution, in addition to a spectral
resolution of a few eV and a much higher effective area. This will
allow to study turbulence also for a significant number of clusters at
much higher redshifts. In addition, the spectral resolution will be
sufficient to study line profiles in detail
\citep[e.g.][]{2003AstL...29..791I}.

\begin{figure}[t]
\begin{center}
\includegraphics[width= 0.9\textwidth]{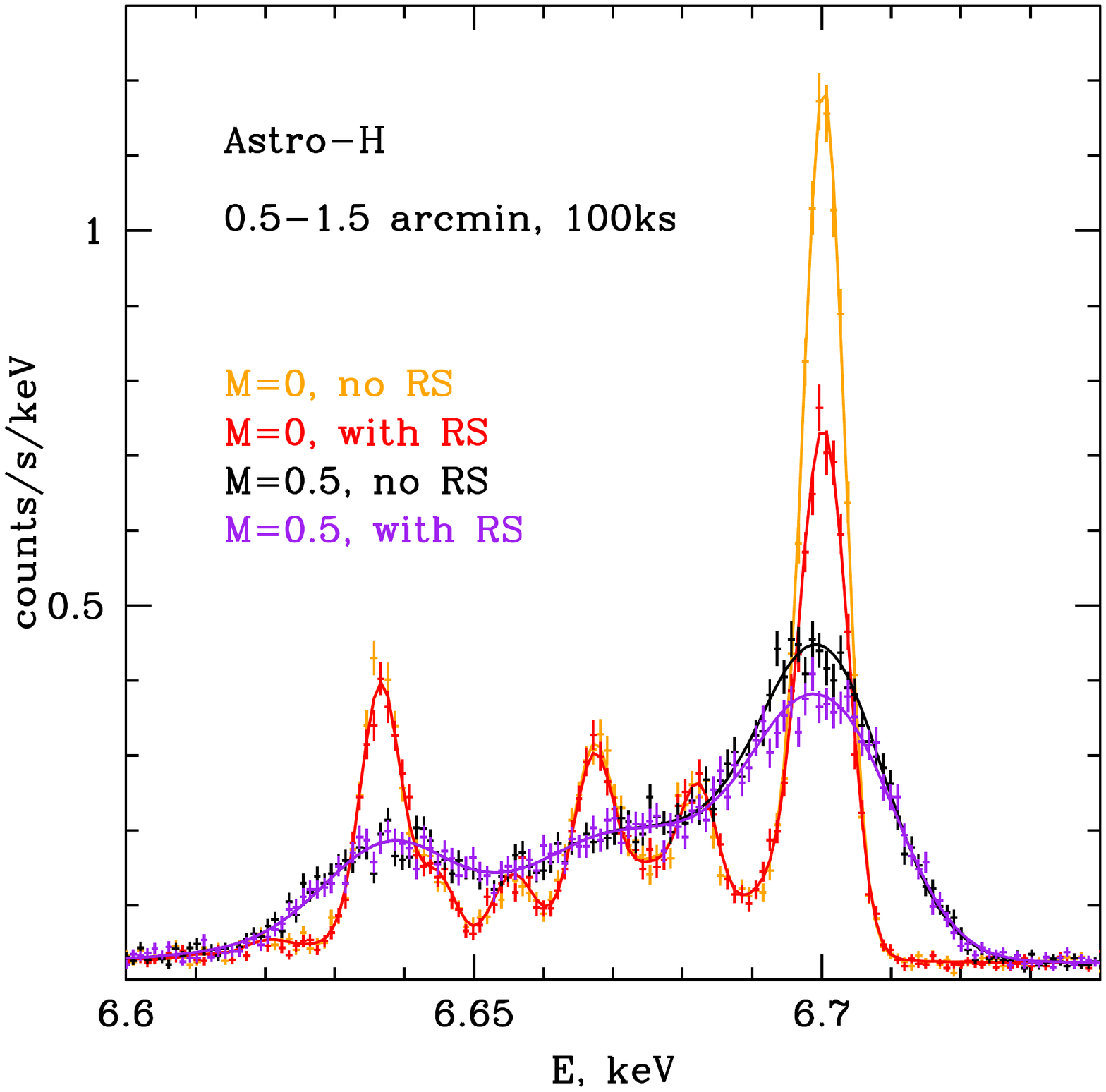}
\end{center}
\caption{Simulated Astro-H spectra from the core (0.5-1.5 arcmin annulus) of the
Perseus Cluster for 100 ks observation.
Points show the simulated spectra with and without resonant
scattering for different levels of turbulence, parametrized through
the effective Mach number ${\cal M}$. Note that 100 ks observation is sufficient
to detect the resonant scattering signal even if gas motions are present.
\citep[Adapted from][]{2013MNRAS.435.3111Z}.}
\label{fig:perseus_rs}
\end{figure}

\subsubsection{Resonance scattering}

Resonance scattering offers a unique tool to study turbulence in
clusters. The hot plasma in clusters of galaxies is in general
optically thin, which means that all radiation that it produces can
freely escape. However, a few of the strongest resonance lines have an
optical depth of the order of unity. Photons emitted in such lines can
be scattered several times before they finally
escape. For clusters of galaxies, the method was pioneered by
\citet{1987SvAL...13....3G}, see \citet{2010SSRv..157..193C} for a
review.  Essentially, line photons from resonance lines emitted from
the center of the cluster towards the observer will be scattered out
of the line of sight, but are not lost. This leads to an apparent
dimming of the line towards the cluster center, but a brightening
towards the outer parts. Because of the higher surface brightness near
the cluster center, this dimming of resonance lines towards the center
is the easiest to observe.

To make fully use of this effect, two lines of the same ion are
needed, with different oscillator strength. Then from the radial
profile of their line ratio (combined with information about the
temperature and density profile of the cluster obtained from spectral
modeling), the optical depth of the scattering line can be
determined. This optical depth depends on the ratio of the turbulent
to thermal energy density in the gas. Therefore this method directly
probes the turbulence. Since the core of the cluster is typically much
brighter than the outskirts, the optical depth is particularly
sensitive to the radial component of the velocity field and
small-scale eddies, while large-scale coherent motions tend to shift
the line energy, but do not affect the optical depth
\citep{2011AstL...37..141Z}.

One of the most promising ions to apply this method is Fe~XXV. The
1s--2p and 1s--3p transitions of this ions have high and low
oscillator strength, respectively. While BeppoSAX and ASCA
observations of the Perseus cluster indicated
 the possible presence of
a strong scattering effect and hence low levels of turbulence \citep{1998ApJ...499..608M,1999AN....320..283A},
subsequent observations with XMM-Newton showed that the Fe~XXV 1s--3p
line is contaminated with nickel line emission; correcting for that
results in insignificant resonance scattering hence the presence of
substantial turbulence, with characteristic speeds of 0.36 times the
sound speed and hence a turbulent pressure of about 10\% of the
thermal pressure
\citep{2004ApJ...600..670G,2004MNRAS.347...29C}. Future ASTRO-H
mission will have sufficient sensitivity and energy resolution to
unambiguously identify signatures of the resonant scattering in the
Fe~XXV triplet (Fig.~\ref{fig:perseus_rs}) in a 100 ks observations.

Another interesting diagnostic ion is Fe XVII. This has a strong
resonance line at 15~\AA, and by comparing this line with other
Fe~XVII lines, \citet{2002ApJ...579..600X} discovered resonance
scattering in the elliptical NGC~4636 using RGS data. This work was
extended by \citet{2009MNRAS.398...23W} using five giant elliptical
haloes, and using in addition to the RGS spectra also high-resolution
images from Chandra. In four of these systems, the 15~\AA\ line is
suppressed relative to the other Fe~XVII lines. For NGC~4636 this
leads to a turbulent velocity of less than 100~km\,s$^{-1}$ or less
than 5\% for the turbulent to total pressure ratio.

\citet{2012A&A...539A..34D} obtained much deeper RGS spectra for two
of the ellipticals studied by
\citet{2009MNRAS.398...23W}. Interestingly, in these two systems the
amount of turbulence appears to be rather high: in NGC~5044 turbulent
pressure constitutes at least 40\% of the total (with turbulent
velocities between 320--720~km\,s$^{-1}$) and in NGC~5813 turbulence
contributes only 15--45\% to the total pressure, with velocities
between 140--540~km\,s$^{-1}$. However, \citet{2012A&A...539A..34D}
also point out that the atomic physics of Fe~XVII is by no means
undisputed, and some of these results depend on the adopted atomic
parameters. Fe~XVII still remains one of the most difficult and
controversial ions, despite its relatively frequent occurrence in
X-ray spectra.

The profiles of spectral lines that are affected by resonance
scattering will be different from those of non-resonant lines. Because
the optical depth in the line core is higher than that in the line
wings, the core will be suppressed relative to the wings
\citep[e.g.][]{1987SvAL...13....3G,2013MNRAS.433.1172S}. The
characteristic depression at the center of the line could therefore be
used as a proxy for resonant scattering when observing bright cluster
cores.

Apart from the distortions of the line intensity and shape, the
resonant scattering causes polarization of the scattered line flux
\citep{2002MNRAS.333..191S} at the level of 10-15\%. The polarization
plane is expected to be perpendicular to the direction towards the
cluster center. Gas motions will affect both the degree and the
direction of polarization \citep{2010MNRAS.403..129Z}. The changes in
the polarization signal are in particular sensitive to the gas motions
perpendicular to the line of sight. This opens a principal possibility
of measuring {\it transverse} component of the velocity field, once
sensitive X-ray polarimeters with good spectral resolution become
available.

\section{Summary and conclusions}
\label{sec:summary}

Between the end of the 90's and the beginning of the
  2000's, the X-ray satellites Chandra and XMM-Newton have definitely proven
  that galaxy clusters are extremely exciting physics laboratory for
  the characterisation of complex plasma processes. In the same years,
  the development of deep radio observations of clusters have
  confirmed the presence of a non-thermal intracluster component
  (cosmic rays and magnetic fields) in the ICM, whose physical
  properties seem to be strongly connected to the dynamical state and
  evolutionary history of clusters. As presented in the previous
  sections, in the next years a new generation of X-ray and radio
  telescopes will allow us to do a further step forward in the
  characterisation of the MHD physical processes that we expect to be
  shaping the ICM properties, but for which we are not yet able to
  make precise measurements.

Great advancements are expected in the direct
  measurements of turbulence thanks to the increased kinematic
  resolution of the X-ray satellites ASTRO-H first and Athena
  then. This will have a relevant impact on both astrophysical and
  cosmological cluster studies. As described in this paper, turbulent
  plasma motions are expected to play an important role in the AGN
  energy dissipation at the center of galaxy clusters and, together
  with merger induced shocks, in the acceleration of intracluster
  cosmic rays and in the amplification of magnetic fields. The
  greatest impact in the characterisation of the non-thermal cluster
  components is expected to come from new and future radio
  interferometers \citep[SKA in particular and, before, its precursors
  and pathfinders, such as JVLA, LOFAR, MWA, LWA, ASKAP, MeerKAT,
  \dots; see][and references therein]{norris13}. All together, these
  instruments will cover a broad spectral range, from a few tens of
  MHz to, possibly, $\sim$15 GHz. Broad-band low-frequency
  observations at arcsec resolution will be key to systematic searches
  of steep-spectrum sources, from radio ghosts at the center of
  clusters, to diffuse radio sources, expected to be hosted in a big
  fraction of low-mass and minor merging clusters
  \citep{2012A&A...548A.100C}. Both total intensity and radio
  polarisation observations from SKA and its pathfinders will allow us
  to access the magnetic field intensity and structure, as well as the
  acceleration mechanisms responsible for electron acceleration,
  within the intra- and, possibly, inter-cluster volume up to $z \sim
  1$ \citep[e.g.][]{bonafede15,ferrari15,govoni15,vazza15}.

Since random turbulent motions are expected to provide
  a pressure support to the ICM (thus affecting the measure of cluster
  masses based on the assumption of hydrostatic equilibrium), a
  precise quantification of the turbulence pressure is not only
  interesting from the astrophysical point of view, but also crucial
  for improving the constraints on cosmological parameters through
  cluster number counts. To this respect, the eROSITA satellite, to be
  launched in 2016, will survey the sky with unprecedented sensitivity
  and is expected to detect about $10^5$ galaxy clusters down to $5
  \times 10^{13} M_{\odot}/h$ and with a median redshift $z \sim 0.35$
  \citep{2012MNRAS.422...44P}\footnote{A flat cosmology with
    $\Omega_{\Lambda} = 1 - \Omega_{m}$ and $h=0.701$ is assumed here,
    see \citet{2012MNRAS.422...44P}.}. Later on, the wide field imager
  onboard of Athena is expected to conduct blind field surveys,
  allowing to detect and characterise the mass and dynamical state of
  clusters and groups up to and beyond $z=1$. To be noted that
  important developments are ongoing also in the (sub-)mm domain,
  allowing deep improvements in the observations of the
  Sunyaev-Zel'dovich effect (SZE)\footnote{The SZE is the change in
    the apparent brightness of the Cosmic Microwave Background
    radiation towards a cluster of galaxies due to inverse Compton
    interaction between CMB photons and intracluster electrons.,
  which provide a complementary and powerful tool for detecting
  clusters and for identifying pressure sub-structures in their
  atmospheres
  \cite[e.g.][]{2011ApJ...734...10K,2013ApJ...763..127R,2013JCAP...07..008H,2014A&A...571A..29P}. }

The gamma-ray observations reviewed above have placed stringent limits
on the pressure contained in protons of energies above 100 MeV with hard energy spectra.
The constraints on the energy density in non-thermal particles with steep enough spectra  dominated by particles
of energies well below 100 MeV remain to be established with the new low energy gamma-ray experiments
which are under development \citep[see e.g.][]{pvb12,lebrun14}.

We can conclude that the excellent synergy between
  future X-ray satellites, together with the huge developments in the
  new generation of radio and mm telescopes, will allow us a
  breakthroughs in our understanding of the evolutionary physics of
  the intracluster plasma and in the exploitation of galaxy clusters
  as tools for cosmology.

\begin{acknowledgements}
  We would like to thank the referee for useful comments and the
  ISSI staff for providing an inspiring atmosphere favourable for intense
  discussions. We thank Hiroki Akamatsu for providing us with Figure ~
  \ref{fig:CIZA} before publication. W. Forman acknowledges support
  from NASA contract NASA-03060 that funds the Chandra HRC project,
  the Chandra archive grant AR1-12007X, and the NASA observing grant
  GO2-13005X.  SRON is financially supported by NWO (the Netherlands
  Organization for Scientific Research).
\end{acknowledgements}


\bibliographystyle{aps-nameyear}
\bibliography{ferrari}                
\nocite{*}


\end{document}